\documentclass[aps,amssymb,amsmath,reprint]{revtex4-1} 

\usepackage{graphicx}  
\usepackage{dcolumn}   
\usepackage{physics}
\usepackage{color}
\usepackage{appendix}
\usepackage[normalem]{ulem}
\usepackage{lineno}
\usepackage{microtype}
\usepackage{upgreek}
\usepackage{xr} 
\setcitestyle{open={(},close={)}}

\definecolor{orange}{rgb}{0.8, 0.3, 0}

\definecolor{blueviolet}{rgb}{0.2, 0.2, 0.6}
\usepackage[pdftex, 
bookmarks=true, 
colorlinks=true,
allcolors=blueviolet,
pdfstartview={FitH}, 
]{hyperref}

\usepackage{comment}

\newcommand{\arxivversion}{}

\begin{document}

\title{Coulomb blockade in microscopic material defects as a source of decoherence and noise in solid-state quantum circuits}

\author{R. Banerjee$^{1}$}
\email{email: riju.banerjee@npl.co.uk}
\author{L. P. Lindoy$^{1}$}
\author{M. Hegedüs$^{1,2}$}
\author{A. Hutcheson$^{1}$}
\author{T. Hawkins$^{1}$}
\author{E. Daghigh-Ahmadi$^{1}$}
\author{S. Samaddar$^{1}$}
\author{T. Barker$^{1}$}
\author{J. P. Goff$^{2}$}
\author{A. Ya. Tzalenchuk$^{1,2}$}
\author{I. Rungger$^{1,3}$}
\author{S. E. de Graaf$^{1}$}
\email{email: sdg@npl.co.uk}

\affiliation{$^1$National Physical Laboratory, Teddington TW11 0LW, United Kingdom}
\affiliation{$^{2}$Physics Department, Royal Holloway University of London, Egham TW20 0EX, United Kingdom } 
\affiliation{$^{3}$Department of Computer Science, Royal Holloway University of London, Egham TW20 0EX, United Kingdom }

\begin{abstract}

A critical limitation of solid-state quantum devices arises from the materials from which they are fabricated: uncontrolled surfaces, interfaces, and structural imperfections introduce numerous sources of loss and decoherence. Despite extensive efforts, linking these decoherence mechanisms to their microscopic material origins---essential for developing effective mitigation strategies---remains an outstanding challenge that has slowed coherence improvements. Here, using scanning gate microscopy on live superconducting circuits we identify a previously unrecognised decoherence mechanism originating from Coulomb blockade and microwave-driven charge tunnelling in metallic grains. Such grains are ubiquitous in thin-film devices fabricated by standard lithography processes. By characterising multiple defects across different devices, we find such defects to be as common and as debilitating to device performance as two-level system (TLS) defects, while originating from a fundamentally different physical mechanism. Importantly, conventional characterisation techniques would misattribute this loss to other, microwave power-independent processes. Our observations thus reveal a widespread source of decoherence in superconducting circuits, challenging the prevailing paradigm that coherence lifetimes are primarily limited by TLS defects. Eliminating metallic grains during fabrication provides a clear and practical route to suppress this mechanism, offering a pathway towards improved coherence and reduced noise in microwave-based solid-state quantum devices.
\end{abstract}
\maketitle

\section*{Introduction}
The physical origins of material defects that cause loss and decoherence remains an open question, slowing down the development of solid-state quantum technologies \cite{Siddiqi2021}. Diverse applications, from quantum sensing to large-scale quantum computers all stand to benefit from reduced noise and improved quantum coherence. For superconducting circuits, decoherence from extrinsic sources such as non-equilibrium quasiparticles generated by high-energy photons \cite{Aumentado2004, Houzet2019}, stray magnetic fields and ionising radiation \cite{Vepsalainen2020, McEwen2022, Cardani2023} can be reduced by filtering and shielding. However, intrinsic materials and fabrication-related decoherence sources have been much harder to eliminate and are currently the primary barrier to reaching the fault-tolerant threshold for quantum computing.  Several recent experiments have focused on in-situ techniques to identify these intrinsic materials-sources of decoherence so that strategies can be developed to eliminate them \cite{Muller2019, degraaf_2017, Hegedus2024, Bilmes_2022, Lucas_2023, Gunzler_2025, deGraaf_2018, Kumar2016}. An alternate approach has
aimed to correlate device performance with structural or chemical properties of intrinsic defects via ex-situ materials science techniques \cite{deGraaf2022, Murthy2025, Altoe_2022, Biznarova_2024, Premkumar_2021, Murthy_2022}. Despite these efforts, the intricate intertwining of the different internal and external decoherence mechanisms \cite{Thorbeck2023, Hirotsuru2025, deGraaf2020b, Spiecker_2023} has made it very challenging to isolate individual defects and determine the mechanisms through which they contribute to circuit loss.

 Here, studying live superconducting circuits with a mK scanning gate microscope (SGM), we discover a new decoherence mechanism arising from Coulomb blockade and charge-tunnelling across spurious metallic islands (grains) within the materials, driven by the microwave field of the device. This clear physical origin and microscopic nature makes mitigation approachable by ex-situ materials science. These defects also exhibit stochastic temporal charge fluctuations, resulting not only in losses but also noise and temporal fluctuations in device dissipation and frequency. Our technique reveals that these defects are seemingly ubiquitous and as detrimental to device performance as two-level system (TLS) defects \cite{Muller2019}. TLS are commonly used as an umbrella term to describe spurious interactions of a device with unknown quantum degrees of freedom present in its environment; here we demonstrate that some defects with similar impacts on device performance as TLS, are in fact of entirely different physical origin.

\begin{figure*}[t!]
\centering
\includegraphics[width=1.0\textwidth]{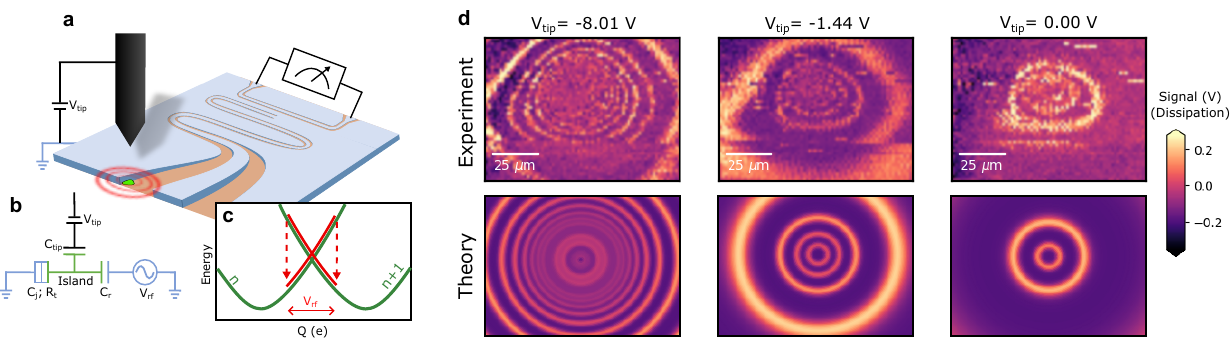}
\caption{{\bf{Coulomb blockade in tunnel-coupled microscopic islands causes dissipation in superconducting quantum circuits.}} (a) Schematic of the experimental setup: A conducting scanning probe tip (black) with voltage $V_{\rm tip}$ is placed above a grounded superconducting resonator, whose $S_{21}$ response at a frequency $f_{\rm readout}$ slightly off resonance is continuously monitored. A microscopic island (green) acts as a defect coupling to the microwave electric fields of the resonator. Sweeping the tip voltage varies the island's electrostatic potential, inducing a change in its charge state. (b) Equivalent circuit of the setup. (c) Microwave voltage $V_{\rm rf}$ from the device drives the system along the energy levels of the defect charge states (green parabolas) along red lines before relaxing (red dashed arrows). (d) The microwave transmission $S_{21} (f_{\rm readout})$ as a function of tip position reveals a series of concentric contours at different tip voltages (top panels). The data were taken with $Z_{\rm tip} = 10\ \mathrm{\upmu m}$. The bottom panels show theoretical simulations (see Simulations section for details) accurately replicating the observations of the top panels. The only parameter varied between the simulation frames is the tip voltage, in accordance with the experiment. }
\end{figure*}

Understanding the mechanisms of charge noise has been a central problem \cite{Paladino_2014} ever since the development of Coulomb blockade-based single charge devices \cite{Averin_1986, Schoelkopf_1998, Nakamura1999}, and is particularly relevant for superconducting circuits. Charge tunnelling in microscopic grains are a source of charge noise in charge qubits \cite{Kafanov2008}, and superconducting qubit architectures have since been engineered to reduce their susceptibility to charge noise \cite{Koch_2007, Manucharyan_2009, Yan_2016}, significantly improving coherence lifetimes. However, such engineering techniques are ineffective against the microwave-driven Coulomb blockade-based mechanism we reveal here, which is analogous to Sisyphus dissipation in engineered rf single-charge devices \cite{Persson_2010, GonzalezZalba_2015}, with the defect--device dynamics described by an Anderson--Holstein-type impurity model.

 Our findings are applicable well beyond superconducting circuits, with implications for a range of cryogenic solid-state devices fabricated from metallic thin films and operated by microwaves.  
 Unintended metallic grains in which this mechanism occurs can form naturally in thin-film devices, especially near oxidised patterned edges. We expect these defects to be particularly prevalent in strongly disordered and granular superconductors \cite{Grunhaupt_2019, Kamenov_2020}, possibly also contributing to the excess dissipation found in them \cite{Charpentier_2026}. 
 However, modern materials science tools are well poised to identify this type of defect at scale, providing an efficient way to steer materials and process development towards improved device coherence through, e.g. optimisation of thin-film deposition conditions \cite{Drimmer_2025}, development of epitaxial thin-film growth \cite{Do_2026, Kline_2011, Weides_2011} and encapsulation techniques \cite{Bal_2024}.

The principle of our experiment is shown in Fig.~1a. We study several different superconducting hanger resonators, making similar observations in all of them. Here we sketch a grounded coplanar-waveguide resonator as a representative example. The resonator couples via its microwave electric field to a randomly located defect (island) represented by a green dot. The SGM setup consists of an etched tungsten tip attached to a quartz tuning fork for atomic force microscopy (AFM) to image and locate topographic features on the sample. The setup is placed inside a magnetically shielded light-tight enclosure housed at the bottom of a dilution refrigerator with a base temperature of 15 mK. A calibrated $\rm RuO_x$ thermometer placed underneath the sample stage reaches a base temperature of $\sim 45$ mK when the SGM is idle. The setup is described in more detail in \textcite{Hegedus2024}. 
The equivalent electrical circuit of the defect, resonator and tip is shown in Fig.~1b. Here $C_\mathrm{j}$, $C_\mathrm{r}$ and $C_\mathrm{tip}$ denote the coupling capacitances of the island to the ground plane, the resonator and the tip, respectively.

To detect defects coupled to the resonator, the microwave transmission signal $S_{21}$ is continuously measured at a frequency $f_\mathrm{readout}$, slightly offset from the resonator frequency $f_\mathrm{res}$ for maximum contrast, using a heterodyne technique. The electrostatic environment of the device is locally tuned by placing the tip above it at a constant height $Z_\mathrm{tip}$ and varying its voltage $V_\mathrm{tip}$. When the tip electric field tunes the defect state, the measured $S_{21} (f_\mathrm{readout})$ changes. The tip is then moved in a grid pattern. At each grid point in the $xy$-plane, the tip voltage is swept and the corresponding $S_{21}(f_\mathrm{readout})$ is recorded for each $V_\mathrm{tip}$ value.

\begin{figure*}[t!]
\centering
\includegraphics[width=0.8\textwidth]{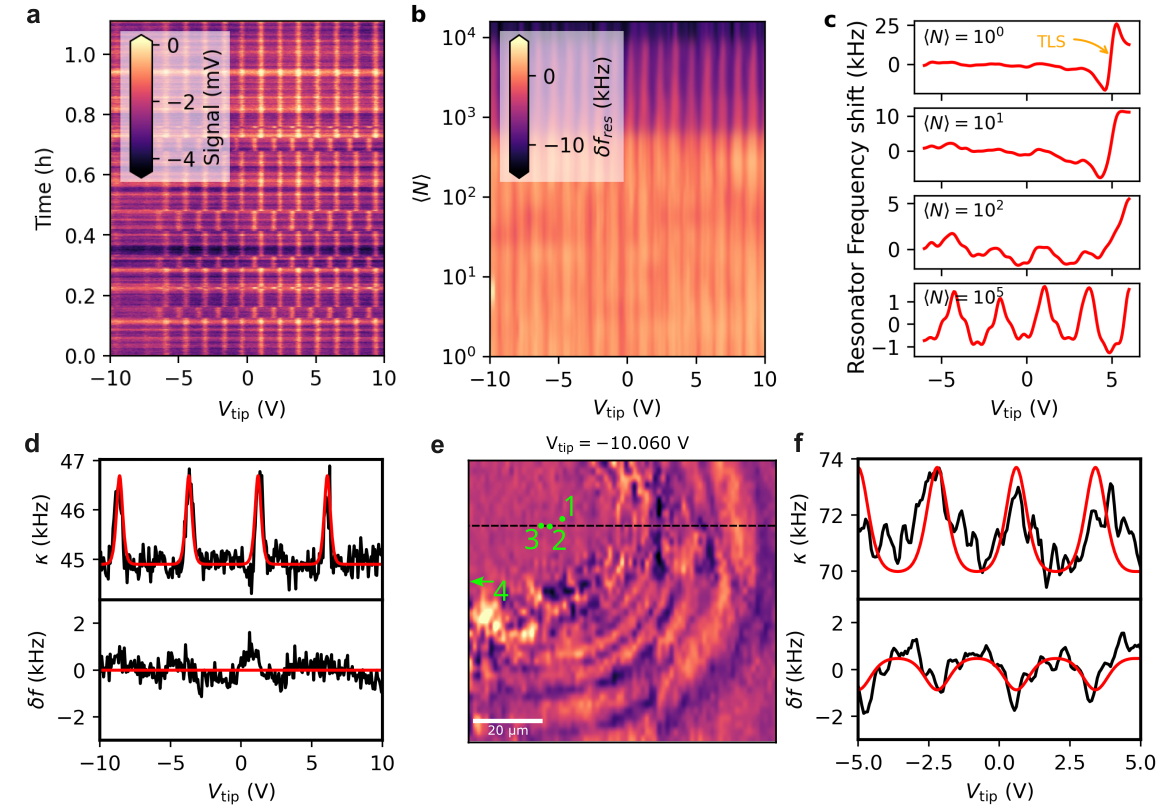}
\caption{{\bf{Quantifying dissipation and frequency shifts.}}
(a) Measured heterodyne response from the resonator when keeping the tip fixed $10\ \upmu\rm m$ above the defect in Fig.~1 and repeatedly ramping $V_{\rm tip}$. A large number of equidistant peaks and stochastic temporal fluctuations of the defect offset charge are observed. Each voltage sweep is 2 seconds long. Defect charge state fluctuations occur on the timescale of minutes. 
(b) Resonator frequency as a function of tip voltage and resonator driving power for a different defect, showing a Sisyphus response unaffected by power. $T=350$ mK, $Z_{\text{tip}}=10\mathrm{\upmu m}$. 
(c) Resonator transmission and the extracted centre frequency as a function of tip voltage for increasing resonator drive powers. $T=440$ mK, $Z_{\text{tip}}=100\mathrm{\upmu m}$. A TLS is present at $+5$ V at this point in time, resulting in an avoided crossing that is saturated at high powers, whereas the periodic Sisyphus response remains unsaturated. Note that $y$-scales of the panels are different. The data were taken in the field of view shown in Fig.~2e.
(d) Measured (black) loss rate and frequency shift (absence of) for the defect in Fig.~1 fitted to simulations (red). Simulation parameters were $f_\mathrm{res} = 5.3$ GHz, $R_\mathrm{t} = 600\ \mathrm{k}\Omega$, $C_\mathrm{j} = 0.1$ fF, $C_\mathrm{r} = 5\times10^{-6}$ fF, $C_\mathrm{tip} = 3.25\times 10^{-5}$ fF. 
(e) Voltage slice of data taken on a Nb on Si sample (resonator C, see SI for details). Dashed line and numbered points refer to the coordinate axis and location of defects identified in Fig.~4.  
(f) The measured resonator loss rate $\kappa$ and frequency shift $\delta f$ (black) for the dominant defect in (e) with fitted simulations (red). Simulation parameters were $f_\mathrm{res} = 6.7$ GHz, $R_\mathrm{t} = 15\ \mathrm{k}\Omega$, $C_\mathrm{j} = 0.35$ fF, $C_\mathrm{r} = 5\times10^{-6}$ fF, $C_\mathrm{tip} = 5.75\times10^{-5}$ fF. Smaller oscillations in the experimental data are due to multiple defects. 
}
\end{figure*}

{\bf{Observations.}} Slices of such grid data $S_{21} (x, y, V_{\rm tip}=\rm constant)$ taken above a NbN resonator patterned on a sapphire substrate (Resonator A) are presented in Fig.~1d (top panels). They show multiple concentric rings that shrink (and then grow) as $V_{\rm tip}$ is swept from negative to positive voltages. A change in the measured $S_{21} (f_{\rm readout})$ signal (colour scale in Fig.~1d) implies a change in the resonator quality factor $Q$ and/or frequency $f_{\rm res}$ in response to the defect. 
A typical voltage spectrum taken at a fixed point in space shows a series of peaks equidistant in $V_{\rm tip}$, shown in Fig.~2a and in more detail also in the Supplementary Information (SI). We have observed over 20 equidistant peaks on either side of $V_{\rm tip}=0$, being limited by the tip voltage that we can safely apply without risking damaging the sample. 
We also observe random offset charge jumps in time, shifting the phase of the periodic response in voltage (Fig.~2a), and an inability to saturate the defect response by readout power (Fig.~2b).
We also show a second dataset in Fig.~2c, taken at a moment in time where a TLS defect is also present and couples strongly to the resonator at $V_{\rm tip}=5$ V, yielding an avoided-crossing and a large frequency shift. Increasing the number of photons in the resonator easily saturates the TLS, while the periodic response remains intact. This observation demonstrates that the mechanism is distinct from TLS. The defects were also found to survive thermal cycling to room temperature, remaining in the same location with the same characteristics.

Taken together, these observations are a clear indication of Coulomb blockade physics and the presence of an electron box. Each ring in Fig.~1d corresponds to a contour of constant charge induced on the defect due to the electric field from the tip.
 This also implies that the defect is of microscopic material origin, likely a metallic grain which happens to be tunnel-coupled to an electronic reservoir and located within the microwave mode volume of the resonator. 
 We also note that conductive-tip AFM has been previously used to study low-frequency mechanical dissipation caused by charge tunnelling in quantum dots, yielding similar concentric rings \cite{Walkup_2023, Miyahara_2017, Roy_2015}.

Regardless of the microscopic origin, a model of an island with discrete charge states tunnel-coupled to a continuum, and driven by the microwave field from the device, fully reproduce the concentric ring pattern, both in terms of ring spacing and lineshapes (Fig.~1d). The system is described by a series of charge states with energy dispersion parabolic in the induced charge $Q$ on the island, 
$E_\mathrm s = (Q - C_\mathrm{tip}V_\mathrm{tip})^2/2C_{\Sigma}$,
where $C_{\Sigma}$ is the sum of all capacitances to the island.
Increasing $ V_\mathrm{tip}$ favours adding an extra charge, taking the system from the $n^\mathrm{th}$ to the $(n+1)^\mathrm{th}$ charge state. Near the charge degeneracy point, the microwave voltage $ V_\mathrm{rf}$ and thermal fluctuations repeatedly drive the system along one charge state (solid red lines in Fig.~1c), with a relaxation rate to the lower energy state that increases with increased level separation. Relaxation (red dashed arrows) is accompanied by tunnelling of a single charge across the junction, which results in energy dissipation in the resonator. The mechanism is that of the Sisyphus resistance \cite{Persson_2010}, and therefore we henceforth refer to these defects as `Sisyphus defects'.

To separate dissipation and resonance frequency shift we again position the tip at a fixed point near the defect in Fig.~1 and record the full resonator response $S_{21}(f)$ for different $V_\mathrm{tip}$. In Fig.~2d we show the resulting  resonator loss rate $\kappa = f_\mathrm{res}/Q$ and the frequency shift $\delta f = f_\mathrm{res}(V_\mathrm{tip}) - f_\mathrm{res}(V_\mathrm{tip}=0)$, extracted from fitting a standard resonance model \cite{McRae2020}.  Periodic peaks corresponding to an increased loss rate are observed, with any change in $\delta f$ being very small and within the noise.

In Figs.~2e and 2f we show the same type of data taken on a different $\lambda/4$ coplanar-waveguide resonator (Resonator C) fabricated in different materials (200 nm Nb on Si) using a different process in a separate cleanroom, as detailed in the SI. For this defect we find, in addition to periodic increases in $\kappa$, a simultaneous lowering of the resonator frequency (Fig.~2f). 
Both defect regimes shown in Figs.~2d and 2f (with or without a frequency shift of the resonator) are in agreement with our physical model (red lines) that jointly fits both measured quantities $\kappa$ and $\delta f$ accurately.

{\bf{Simulations.}} We next present the details of our model and explore the parameter space for Sisyphus defects in devices, with details presented in the SI. The system is described by an Anderson--Holstein-type Hamiltonian and the coupled resonator--island dynamics are modelled using the Orthodox theory for single electron tunnelling \cite{AverinLikharev1991, 752518} with a mean-field approximation to the coupled dynamics, leading to a set of coupled ODEs for the island charge state populations and resonator variables. We model the resonator as capacitively coupled to a metallic island with discrete charge states, which is tunnel-coupled to a continuum (reservoir), as illustrated in Fig.~1b. The electrostatic energy of the island is given by \cite{Persson_2010}
\begin{equation}
\begin{split}
\hat{H}_\mathrm c &= E_\mathrm c \left(\hat{n} - [n_\mathrm g^0 + n_\mathrm g^{\rm tip} + \hat{n}_\mathrm g^\mathrm r]\right)^2 \\
 & = E_\mathrm c \left(\hat{n}\! - \!\left[\frac{C_\mathrm j V_\mathrm j}{e} + \frac{C_\mathrm{tip} V_\mathrm{tip}}{e} + \frac{C_\mathrm r \hat{V}_\mathrm{rf}}{e}\right]\right)^2 ,\label{eq:dot_hamiltonian}
\end{split}
\end{equation}
where $\hat{n}$ is the number operator for the charge states, $E_\mathrm{c} = e^2/2C_\Sigma$ is the charging energy 
and the offset charge contains external ($n_\mathrm g^0$), SGM tip ($n_\mathrm g^\mathrm{tip}$), and resonator ($\hat{n}_\mathrm g^r$) contributions. 
This final contribution leads to island charge state energies dependent on the state of the resonator, and a change of the island charge state following tunnelling perturbs the resonator dynamics which, to first order, induces resonance frequency shifts and dissipation.

We can derive approximate analytical expressions for the frequency shift and loss rate in the linear response regime to which we fit our data:
$\delta\omega = -A\Gamma^2/2(\Gamma^2+\omega_0^2)$ and $\delta\kappa = A\omega_0\Gamma/(\Gamma^2+\omega_0^2)$ with coupling prefactor $A=(ZE_\mathrm{c}^2C_\mathrm{r}^2\omega_0^2/k_\mathrm{B}Te^2)\sech^2{(\Delta E_0/2k_\mathrm{B}T)}$ and tunnelling rate $\Gamma = \frac{\Delta E_0 R_\mathrm{K}}{hR_\mathrm{t}}\coth(\Delta E_0/2k_\mathrm{B}T)$. Here $\Delta E_0$ is the energy level difference, $Z$ the resonator impedance and $\omega_0 = 2\pi f_{\rm res}$.
This yields a ratio $\delta\omega/\delta\kappa = -\Gamma/2\omega_0 = -\frac{R_\mathrm{K}}{R_\mathrm t} \frac{k_\mathrm{B}T}{h\omega_0}$ at the charge degeneracy $\Delta E_0=0$.

This behaviour originates in the restoring force on the resonator from the island that is driven by the resonator dynamics. $\delta\omega$ arises from the dispersive (in-phase) response of the defect and saturates in the adiabatic regime ($\Gamma\gg\omega_0$) whereas $\delta\kappa$ arises from the out-of-phase response and is maximised for $\Gamma\sim\omega_0$. This leads to the $R_\mathrm t\propto f_{\rm res}^{-1}$ scaling shown in the Fig.~3a inset, implying different devices will sample different parts of the distribution of defect $R_\mathrm t$.

In the slow tunnelling limit ($\Gamma\ll\omega_0$) the defect cannot respond within a drive cycle, leading to suppressed back-action. This is exemplified in Fig.~3, which shows the dissipation and frequency shift of a $f_\mathrm{res}=6.7$ GHz resonator (as in Fig.~2f) across the parameter space of possible $R_\mathrm t$ and $C_\mathrm j$.

For defects with strong coupling ($R_\mathrm{t}\lesssim h/e^2$) the Orthodox treatment is no longer strictly valid as phase fluctuations become significant, but such junctions can still show a non-vanishing Coulomb gap. This leads to a renormalisation of the junction capacitance and an effective temperature scale due to lifetime broadening \cite{Chouvaev_Kuzmin_Golubev_Zaikin_1999}. 
For strongly coupled defects this also implies that the width does not vanish as $T\rightarrow 0$ as temperature-independent mechanisms determine the broadening, consistent with our observed lack of temperature dependence below $T=1$ K. This has important consequences for quantum circuits where these defects are present, as further cooling will not mitigate this decoherence mechanism.

\begin{figure}
\centering
\includegraphics[width=0.5\textwidth]{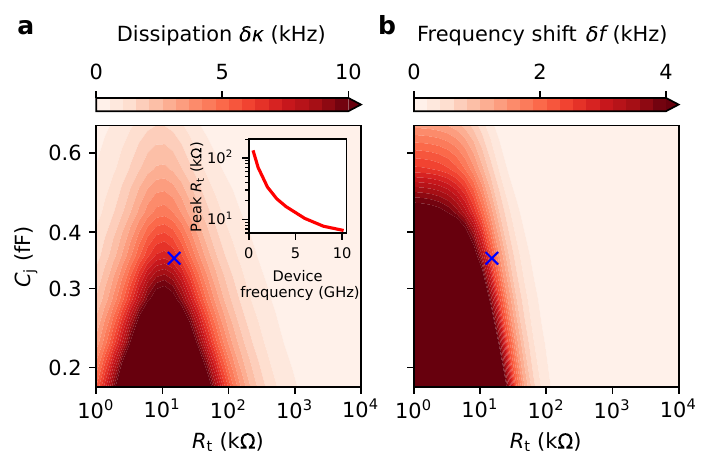}
\caption{{\bf{Decoherence parameter space.}} 
Theoretical numerical evaluation of the expected (a) peak loss rate and (b) frequency shift as a function of defect tunnel resistance $R_\mathrm t$ and tunnel junction capacitance $C_\mathrm{j}$ for a device with $f_\mathrm{res}=6.7$ GHz. The inset in (a) shows the tunnelling resistance $R_\mathrm t$ at which the maximum dissipation occurs as a function of device frequency. The cross indicates the location in the parameter space for defect 3 detected and fitted to theory in Fig.~2e--f.}
\end{figure}

The observed power independence of both $\delta \omega$ and $\delta\kappa$ indicates that the system operates in the linear-response regime. In this limit, the resonator-induced modulation of the island energy remains small compared to the relevant energy scale governing charge fluctuations, such that the tunnelling rates can be expanded to first order in the drive amplitude. As a result, the induced charge response—and hence both the dispersive and dissipative components—scale linearly with the drive. This justifies the use of a linear-response treatment and allows the tunnelling dynamics to be characterised solely through the susceptibility evaluated at the resonator frequency.

As the perturbation is linear in the resonator oscillation amplitude, the resulting back-action force scales proportionally with its oscillation amplitude, leading to the induced frequency shift and damping being independent of photon number. 
We also note that the strength of this effect scales quadratically with single photon voltage amplitude $V_\mathrm{rf}^2 = \hbar\omega_0/2C_r$, such that grains close to resonator metal edges where the microwave electric field is enhanced couple more strongly.
 Consequently, device frequencies, zero-point microwave electric field strength, inter-grain conductance and grain size are accessible experimental parameters which can alter the amount of Sisyphus dissipation in a device.

{\bf{Ex-situ materials science \& grain size.}} Multiple ex-situ techniques verify the presence of grains in the sample with size commensurate with the detected Sisyphus defects. X-ray reflectometry (XRR) performed on a sample from the same Nb on Si wafer in which the defect shown in Fig.~2e was found reveals a 7 nm thick surface oxide layer, and grazing-incidence X-ray diffraction (GIXRD) suggests that the oxide layer is amorphous, and gives an average grain size of 21 nm for the polycrystalline Nb thin film. Scanning Electron Microscopy (SEM) and high-resolution AFM reveals similarly sized grains present in the film and also near patterned edges of the superconducting Nb film.

The change in periodicity with tip--defect distance can be used to estimate the size of the island by measuring the tip--defect capacitance $C_{\rm tip} = q/\Delta V_{\rm tip}$ at different tip locations. Electrostatic simulations in the SI show that all the Sisyphus defects found are consistent with grains of size 10--30 nm residing near the sample surface, and the junction capacitances and resistances extracted from the theoretical modelling are commensurate with such grain sizes (see SI for further details).

\begin{figure}[t!]
\centering
\includegraphics[width=0.5\textwidth]{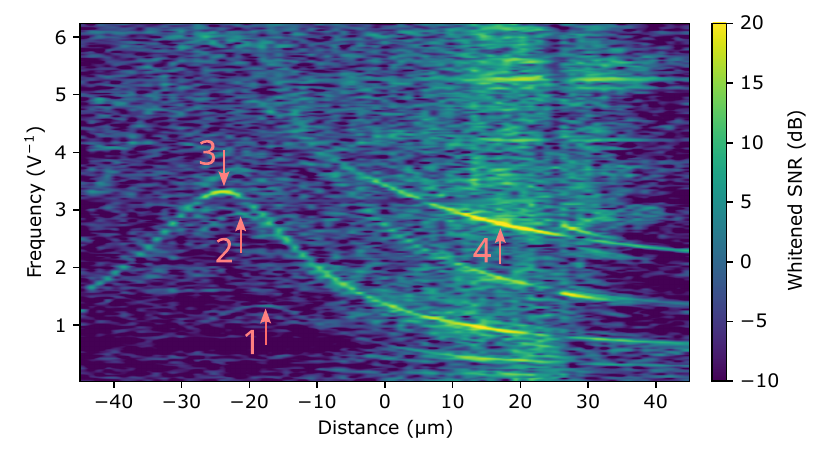}
\caption{{\bf{Abundance of defects.}}
Spectrogram showing Fourier decomposition of the dataset in Fig.~2e along the black dashed line. See text for explanations. We identify four distinct defects (green dots in Fig.~2e). The concentric ring pattern in Fig.~2e is dominated by defect 3, with defect 4 residing outside of the frame. 
}
\end{figure}

{\bf{Abundance.}} Careful inspection of Fig.~2f reveals what appears to be a second periodic variation of $\kappa$ and $\delta f$ with a smaller amplitude but faster period of oscillation. This is actually the superimposed signal from another Sisyphus defect within the same scan frame. 

Using Fourier analysis we are able to identify at least four defects in the dataset shown in Fig.~2e (marked 1--4). As the tip--defect distance is varied, the voltage periodicity $\Delta V_{\rm tip}$ observed in the $S_{21}$ spectra changes, with the frequency of peaks increasing when the tip is closer to the defect. Individual defects located at different positions can be identified by taking the Fourier transform of $S_{21}(V_\mathrm{tip})$ at each tip location along the dashed line in Fig.~2e and producing the spectrogram in Fig.~4. The maximum frequency points marked by arrows 1--3 denote the points along the dashed line in Fig.~2e where the tip--defect distance is minimised (the 4th defect being outside the frame). As the maxima are located at different coordinates along the line, this implies they belong to different defects. More details of the Fourier analysis are provided in the SI, where we also present data from a third resonator revealing at least two additional defects within a small scan region. 

Remarkably, we have only imaged small regions of each of three different resonators, always finding Sisyphus defects, suggesting that they are very common. Our Fourier technique is also only able to detect a small subset of defects that fall within the Nyquist limit and couple strongly enough to the device to yield a detectable change in the loss rate. There are likely many more defects with weaker coupling, that add up to form the total coherence budget.

{\bf{Discussion.}} The deduced grain sizes of 10--30 nm are consistent with previous studies on Nb thin films sputter-deposited under similar conditions, where Nb grains were found to be separated by partially insulating inter-granular regions \cite{Bose_2006}. The granular nature of the film was characterised by a reduced critical temperature, also consistent with our films where $T_\mathrm c = 8.95 $ K $<T_\mathrm{c,bulk}=9.4$ K. Similar films were also found to exhibit power-independent resonator loss, possibly related to grain boundary structure \cite{Drimmer_2025}. 
Nb is known to form an amorphous oxide which can hold closely-packed nm-sized metallic grains in an otherwise insulating matrix \cite{Oh_2023}. Other precipitates, such as hydrides, may also form clusters of similar size \cite{Sung_2025}. A metallic grain separated by an oxide barrier of the right junction transparency towards the rest of the film would create a Sisyphus defect.
We note that in our theoretical modelling and analysis above we have used $1e$ periodicity, but we do not expect any significant differences in a fully superconducting system. The $1e$ parity is consistent with our observed (lack of) temperature dependence and with other single-electron experiments in Nb devices, due to the specific material properties of Nb \cite{Savin_2007}.

Sisyphus defects are naturally occurring and not formed while scanning with the tip, as the defect shown in Figs.~1d and 2d were taken at a location that was not scanned using AFM beforehand. The prevalence of loosely connected grains is naturally expected to be higher along patterned edges of the film, where re-deposits and other contaminants are also more likely. This, combined with the fact that the device's microwave electric field is strongest near the patterned edges, means that it is not unexpected to find multiple defects on a line that corresponds to a geometrical edge in the resonator (Fig.~2c). Additional data in the SI also reveal more defects located near edges of the film. We note that due to the large scan range of our microscope we have significant hysteresis in the tip movement, which limits our accuracy in determining the location of defects to about $1$ $\upmu$m.

Apart from physical grains, other sources of electron boxes are possible, for example dopant clusters \cite{Prabhudesai_2019}, electronic disorder in superconductors \cite{Gantmakher_Dolgopolov_2010}, or processing-induced contaminants. However, our measurements show that the Sisyphus defects are unchanged by thermal cycling to room temperature and extracted capacitances consistent with grain size suggests that in our case physical grains are the most likely origin of the Sisyphus defects found.

We finally quantify the impact of Sisyphus defects on device coherence. 
The relevant quantity that determines if a particular defect causes additional dissipation is given by the peak width compared to the charge periodicity. We find this ratio to be on the order of $10\%$ for the defects observed, and limited by lifetime broadening or external charge noise rather than temperature, typical for electron box experiments at mK temperatures. This `on/off' ratio is independent of tip location, and hence also remains the same in the absence of a tip, i.e. the situation for typical qubit operation. Furthermore, all defects would be subject to their local environment which yields a random offset charge, and this offset charge can also fluctuate in time (Fig.~2a), increasing the likelihood of the defect being `on' at some point in time. Reducing this probability would require methods to suppress the environmental charge noise.

The observed loss rate (up to $\sim 5$ kHz) and the high probability  of interacting with the device means only a handful of defects are required to, on average, limit device coherence to ms timescales. Likely many more defects with weaker coupling to the microwave field exist, that we are unable to resolve in our experiments. Due to its power-independent nature, the Sisyphus loss could easily be misinterpreted as loss due to, e.g. infrared radiation-generated non-equilibrium quasiparticles, magnetic vortex movement or radiation loss in measurements \cite{McRae2020, Alexander_2025}. Comparing Sisyphus defects to TLS, the probability that a TLS is resonant with the device is many orders of magnitude smaller, and only a very small subset of all TLS present will contribute to device loss. Furthermore, a particularly lossy TLS can be detuned by voltage gates to improve coherence, a strategy that does not work for Sisyphus defects.

{\bf{Acknowledgements:}} We thank Gokhan Bakan and Chelsey Zhang for supporting transport measurements, K. Mingard for assistance with the EDS analysis and Nicholas Nugent and Martin Weides for providing the Nb films used. We thank S. Kubatkin, A. Danilov and S. Mahashabde for providing the NbN sample and for helpful discussions. Samples were fabricated in the SuperFab cleanroom at Royal Holloway University of London and at the Chalmers MC2 cleanroom. We acknowledge support from the Engineering and Physical Sciences Research Council (EPSRC) (Grant Number EP/W027526/1), the Horizon Europe European Metrology Partnership project 23FUN08 MetSuperQ where NPL is supported by UKRI grant number 10133632, the UK Department for Science Innovation and Technology through the UK National Quantum Technologies Programme, and Google Faculty Research Awards.



\bibliography{main}
\pagebreak

\ifdefined\arxivversion

\ifdefined\arxivversion

\setcounter{equation}{0}
\setcounter{figure}{0}
\setcounter{table}{0}
\renewcommand{\theequation}{S\arabic{equation}}
\renewcommand{\thefigure}{S\arabic{figure}}
\renewcommand{\thetable}{S\arabic{table}}
\clearpage \begin{centering}{\large{\bf{SUPPLEMENTARY INFORMATION}\\}}\end{centering}\vspace{3mm}

{\let\clearpage\relax \tableofcontents}

\else
\documentclass[aps,amssymb,amsmath,reprint, twocolumn]{revtex4-1}

\fi


\section{Experimental setup and methods}
The SGM microscope operating at mK temperatures is housed inside a Bluefors LD400 dilution refrigerator. The SGM microscope is suspended from springs below the mixing chamber plate of the dilution refrigerator. The sample stage is mounted atop a stack of Attocube steppers used for coarse positioning and piezo scanners in a home-built enclosure. The sample platform contains a RuOx thermometer and a heater, and a PCB around the sample itself brings in the required microwave signals used to measure the sample. 

An etched tungsten tip glued to a quartz tuning fork is used for atomic force microscopy (AFM) imaging to extract the topography and location of the tip above the sample surface. The conductive tip is connected to a voltage source, and the superconducting film on the sample (and the resonators on the sample) is held at ground potential. In SGM mode, we lift the tip a fixed distance above the surface, typically $\sim 10$ $\upmu$m, and sweep the voltage applied to the tip, while moving it above the surface. While sweeping the tip voltage, we measure the $S_{21}$ transmission of the superconducting resonators through on-chip transmission lines.

Two types of measurements are presented in the paper: 

1) For taking grid data, the likes of which are presented in Figs. 1d and 2e of the main text, we use a fast heterodyne technique where we only measure the transmitted microwave signal at a single frequency $f_{\rm readout}$, chosen to be on the shoulder of the resonance lineshape for maximal contrast (see \autoref{fig:Chalmers_res1_q}a for an example). Any change in $f_\mathrm{res}$ or $Q$ will then result in a change in the transmitted signal, which is detected. The frequency $f_{\rm readout}$ is generated from two phase-shifted 30 MHz tones, which are up-converted by combining with a high-frequency tone from a GHz-range local oscillator using an IQ mixer calibrated to minimise spurious sideband signals. The input line also has a variable attenuator at room temperature to vary the power reaching the sample. Output from the fridge is amplified at room temperature to the required level before it is down-converted to 30 MHz and demodulated by a lock-in amplifier before being sent to the Nanonis scanning probe microscope controller. Data from both quadratures are recorded, but for simplicity, here we only show data for one of the quadratures in the manuscript.

2) To detect the loss and frequency shifts of the resonator, we use a different technique where the tip is parked at a spot above and near the resonator, and its voltage $V_{\rm tip}$ is swept. For each $V_{\rm tip}$, full resonance lineshapes ($S_{21} (f)$) are measured by sweeping the heterodyne frequency.  By fitting these lineshapes, the resonator quality factor $Q$ and centre frequency $f_\mathrm{res}$ are then extracted. This technique is applied to extract resonator loss rate $\kappa$ from the quality factor $Q$, and $f_\mathrm{res}$, as in Figs. 2d and 2f of the main text.

\autoref{fig:setupschematic} presents the wiring setup inside the fridge. The signal passes down a heavily attenuated coaxial line to the sample placed inside the microscope. The signal returning from the sample is amplified by a Silent-Waves travelling wave parametric amplifier (TWPA), which is driven by a pump tone at $f_\mathrm{P} \approx 6$ GHz. The signal is then passed through two isolators in series before being further amplified at the 4K stage by a high electron mobility transistor (HEMT) before leaving the fridge. At the mixing chamber stage, we filter the lines using both microwave band-pass filters and commercially available infrared ecosorb filters.

When idle, the SGM sample stage reaches a temperature of $\sim 45$ mK. The sample temperature increases when moving the piezo scanners due to non-negligible heat dissipation. To minimise the impact of heating when grids are obtained, all tip movements are done slowly enough to always keep the temperature of the sample below 1 K, almost an order of magnitude below the critical temperature of our Nb films. Throughout individual measurements and scans, we adjusted the speed to maintain the temperature below 350 mK, unless otherwise stated.

\begin{figure}
    \centering
    \includegraphics[width=0.9\linewidth]{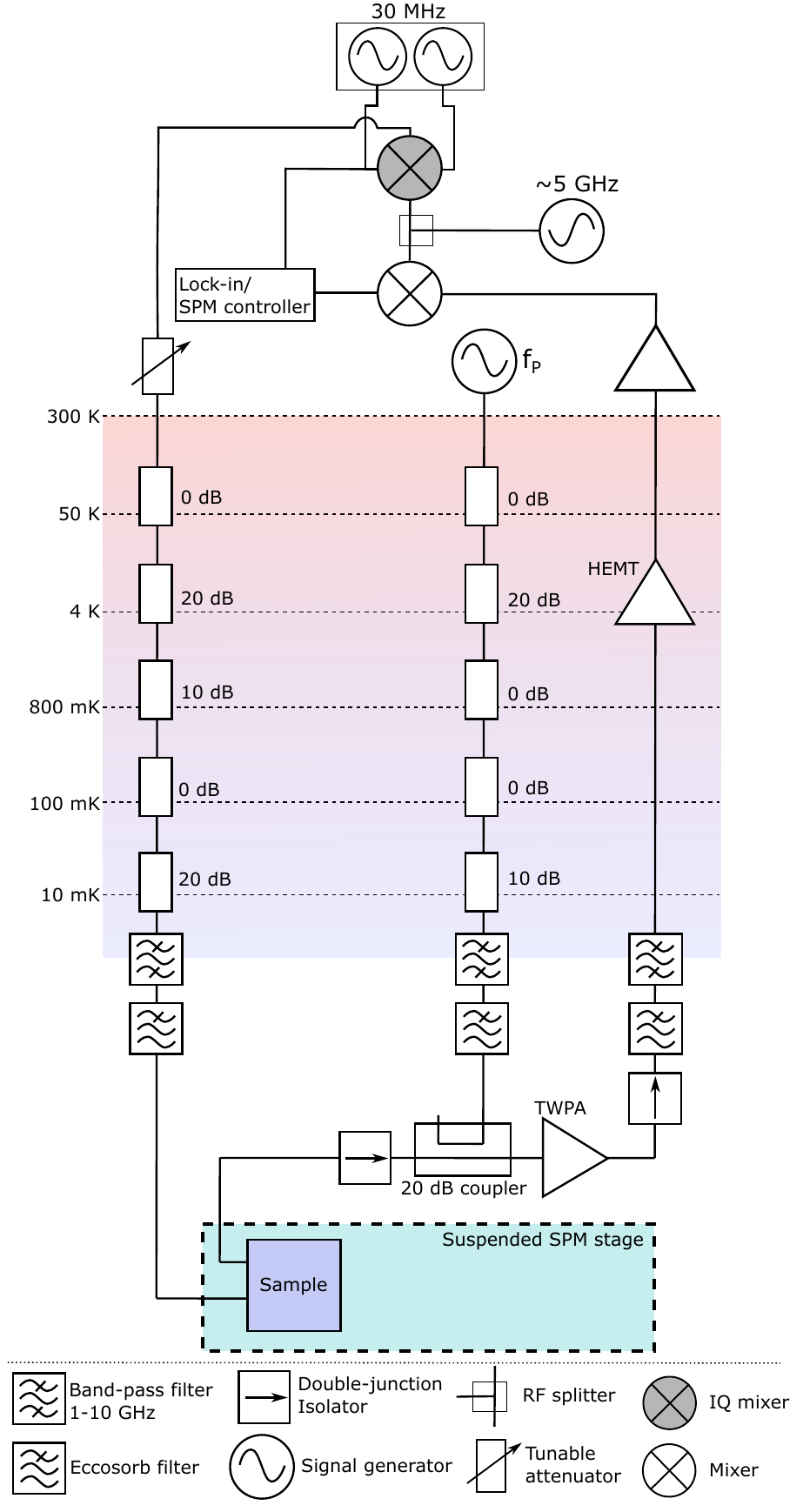}
    \caption{Experimental microwave setup for the cryogenic SGM microscope, showing the wiring configuration inside the dilution refrigerator and the heterodyne measurement setup used at room temperature. }
    \label{fig:setupschematic}
\end{figure}

\section{Imaged devices}

\subsection{Sample 1 -- NbN on Sapphire (Resonator A)}

The data presented in Fig.~1d of the main text were taken on a 40 nm thick $3 \lambda/4$ NbN on sapphire resonator shown in \autoref{fig:Chalmers_res}. The two prongs of the resonator connect at the bottom, and the structure is filled with an interdigitated capacitor. The top of the resonator is open and couples capacitively to the ground and the transmission line. The resonator has a frequency of 5.32 GHz and single photon internal quality factors of $2\cdot 10^5$, as shown in \autoref{fig:Chalmers_res1_q}.

\begin{figure}
    \centering
    \includegraphics[width=1.0\linewidth]{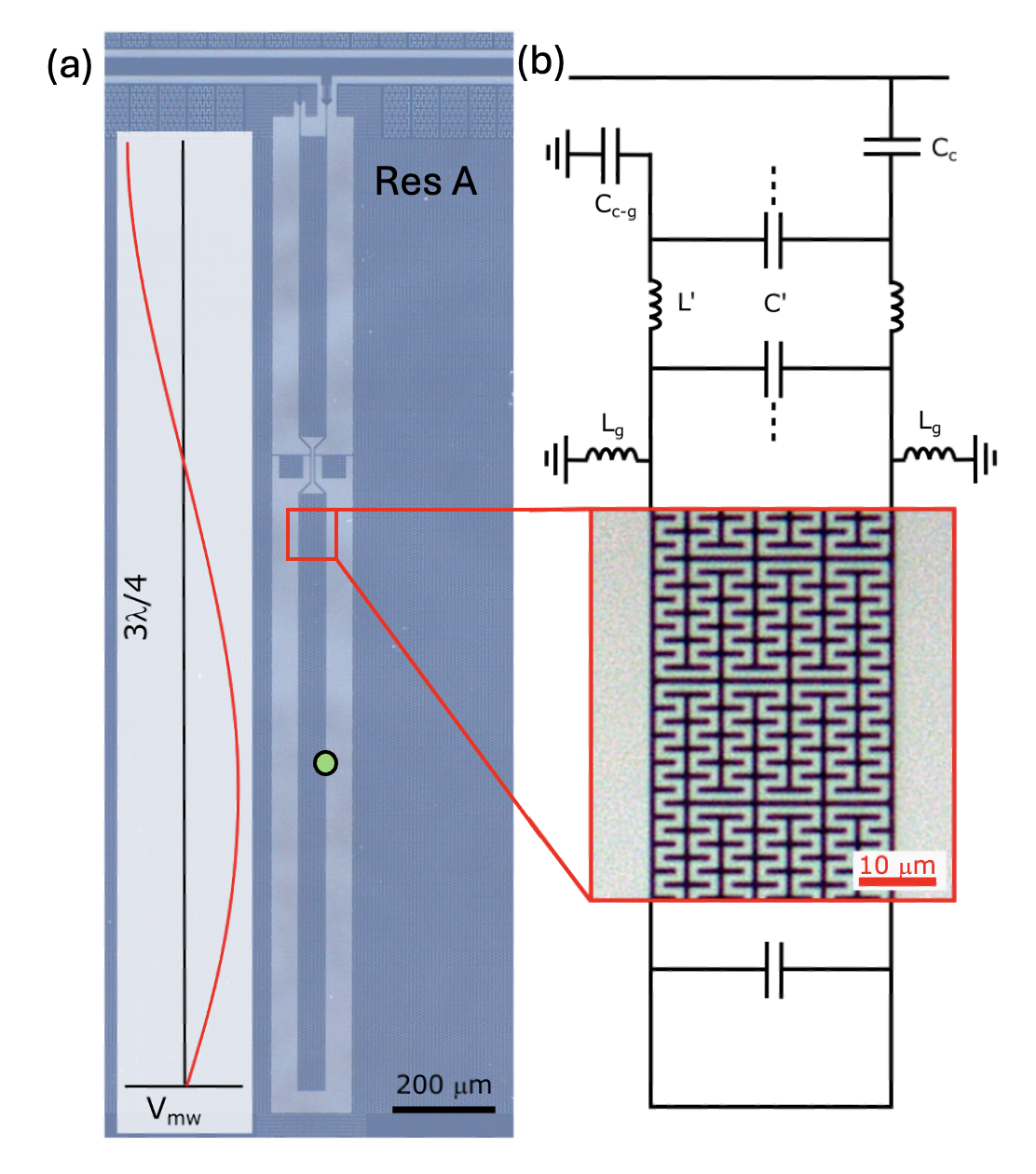}
    \caption{(a) Optical microscope image of the $3 \lambda/4$ NbN on sapphire resonator. The inset plots the profile of the electric field amplitude along the length of the resonator. The green dot corresponds to the location where the Sisyphus defect shown in Fig 1d of the main text was located. (b) shows the equivalent circuit representation of the resonator, with a capacitance per unit length $C^\prime$. The resonator couples to the transmission line and the ground via capacitors $C_\mathrm{c}$ and $C_\mathrm{c-g}$ respectively. Each prong of the resonator also galvanically connects to the ground via inductors $L_\mathrm{g}$, ensuring that the entire structure is grounded. The inset shows a zoomed-in optical image of the interdigitated capacitor forming the resonator, where light corresponds to the sapphire substrate and dark to the NbN thin film. }
    \label{fig:Chalmers_res}
\end{figure}

\begin{figure}
    \centering
    \includegraphics[width=1.0\linewidth]{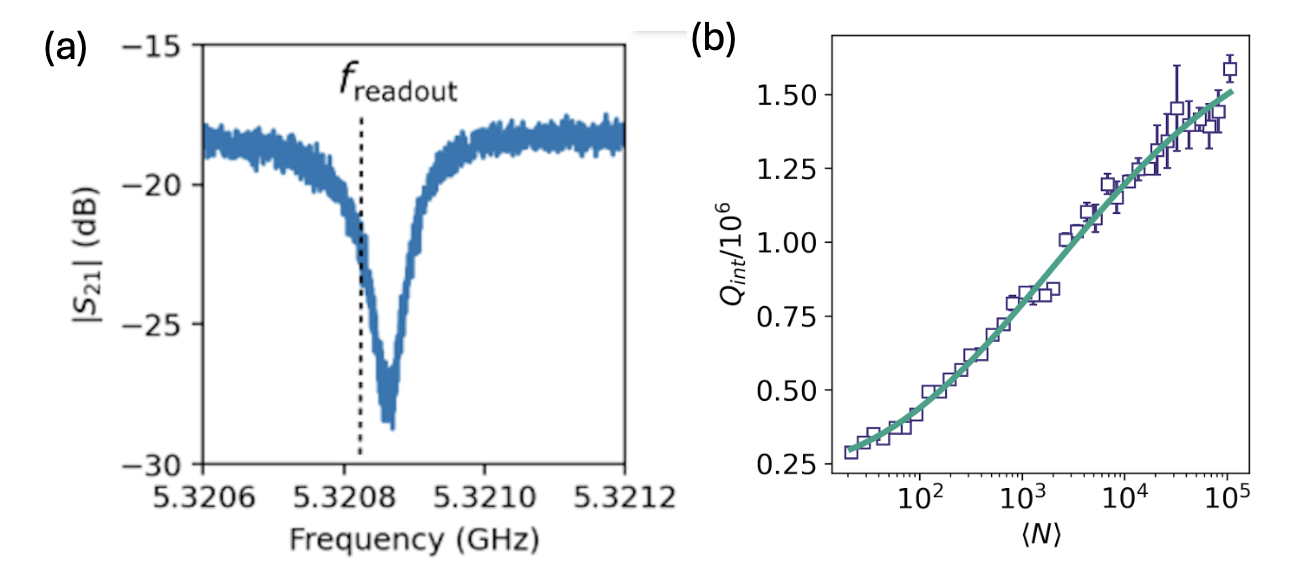}
    \caption{(a) The resonance lineshape of the NbN on sapphire resonator shown in \autoref{fig:Chalmers_res} as measured in our SGM setup, with the variation of its internal quality factor with phonon number in (b).}
    \label{fig:Chalmers_res1_q}
\end{figure}

The sample was fabricated at Chalmers MC2 cleanroom by sputtering a 140 nm thick NbN film on a c-plane sapphire substrate and patterned using electron beam lithography (EBL) and etched in an Ar:Cl$_2$ reactive ion plasma. An additional EBL exposure and etching step performed only on the resonator (avoiding the ground planes) thinned the resonator down to a thickness of 40 nm. This thickness yields the desired sheet inductance to reduce the resonance frequency and match its grounding points to the resonator's voltage node. The ground plane is kept at 140 nm to minimise the impact of spurious ground plane modes arising from a higher sheet inductance. 

 We imaged an area covering $\approx 400$ $\upmu$m along the length of resonator 1 and found one Sisyphus defect in this region, located near the electric field maximum and denoted by the green dot.

\begin{figure*}
    \centering
    \includegraphics[width=0.8\linewidth]{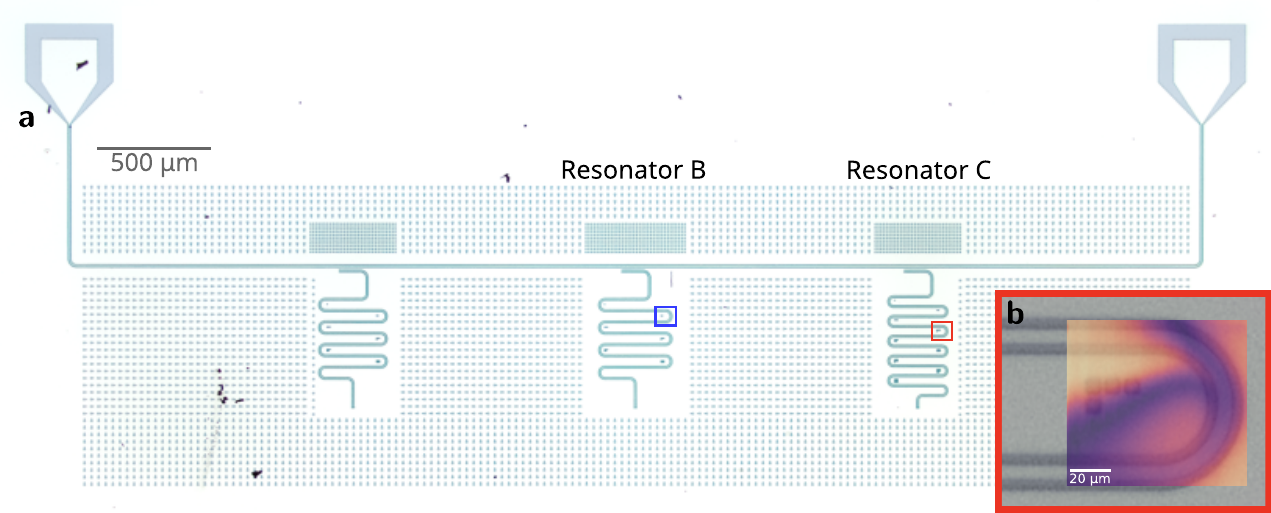}
    \caption{(a) Optical microscope image of the Nb on Si sample with three hanger resonators. The two resonators that were studied are denoted B and C, with the scan regions containing the defects discussed in this paper highlighted. Data from resonator C are presented in Figs. 2c, 2e and 2f of the main text. At least two and four spatially distinct defects were found at these sites, respectively. (b) A close-up of the studied region for resonator C with the capacitive background from the SGM data overlaid, showing a good fit to the resonator geometry. The capacitive signal arises as the tip placed $Z_\mathrm{tip} = 10\ \mathrm{\upmu m}$ above the sample in the resonator's microwave mode volume adds an additional capacitive contribution to the resonator, thereby changing its resonance frequency, detected through the measured transmission $S_{21}(f_\mathrm{readout})$.}
    \label{fig:markersample}
\end{figure*}

\subsection{Sample 2 -- Nb on Si (Resonators B \& C)}
The sample was fabricated from a 200 nm thick Nb thin film on an intrinsic $\langle100\rangle$ silicon substrate, which was cleaned and HF treated before loading into the deposition chamber. This sample was sputter deposited in a DC magnetron sputterer at the cleanroom at the University of Glasgow using a DC power of 207 W and an Argon pressure of 3.7 mTorr. The wafer was kept at room temperature, and sputtering was carried out in 6 steps of 4 minutes each with pauses between steps to allow the wafer to cool. Subsequent patterning forming the device features was done at the SuperFab cleanroom at Royal Holloway University of London, using electron beam lithography and a SF$_6$ inductively coupled plasma (ICP) etch. The etch time was extended to produce deeper $\sim 1.5$ $\upmu$m trenches for easier scanning and locating of device features with mK AFM.

The sample consists of three $\lambda/4$ coplanar waveguide hanger resonators with a centre strip width of 8 $\upmu$m and a gap between centre strip and ground of 5 $\upmu$m. All resonators are open at the top near the transmission line and grounded on the other end. As the walking range of the microscope was limited, in this sample, we imaged and studied two out of three resonators, denoted as Resonator B and Resonator C in \autoref{fig:markersample}. Two Sisyphus defects were found coupled to resonator B (shown in \autoref{fig:SIresB_spectrogram}), and at least four were found coupled to resonator C (shown in Figs. 2e and 4 of the main text and explained later in the Supplementary Information). Resonator C has a length of $L=4.9$ mm, and the Sisyphus defects coupled to it were located at $x=2.7$ mm from the top, where the voltage maximum is located. From this, we estimate the single photon microwave voltage amplitude at the defects' locations to be $V = \cos{(2\pi x/4L)}\sqrt{4 \pi hf_{\mathrm{res}}^2Z} = 2.7\ \mathrm{\upmu V}$ per photon using $Z= 50$ $\Omega$.

\section{Defect properties \& additional data}

\subsection{Typical spectrum}
In \autoref{fig:TypicalSpectra}, we show a typical $S_{21}(f_{\rm readout})$ spectrum obtained when the tip is located at a fixed point near a defect, and its voltage is swept. The tip was $10\ \mathrm{\upmu m}$ above the sample surface (Sample 1). We have observed more than 20 peaks corresponding to different charge states, with the number of peaks depending on the proximity of the tip to the defect and the range of the voltage sweep. The maximum electric field between the tip and the grounded sample was limited to $\pm 2 \times 10^7$ V/m to not damage the sample surface. When multiple defects are present, we see multiple periodicities in the spectra, which can be deciphered by taking a Fourier transform, as in Fig.~4 of the main text.

\autoref{fig:SpectraHeight} shows the variation of the $S_{21}(f_{\rm readout})$ spectra when the tip is kept at a fixed point near the Sisyphus defect shown in Fig.~1 of the main text and moved in the Z-direction only. Increasing the tip-defect distance reduces the electric field of the tip at the defect location, resulting in the peaks spreading out in voltage. 

\begin{figure}
    \centering
    \includegraphics[width=0.9\linewidth]{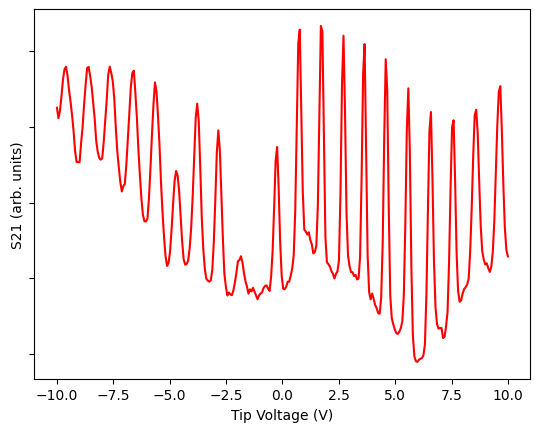}
    \caption{A typical spectrum of a Sisyphus defect shows a series of equally spaced peaks. The data were taken for the defect shown in Fig.~1 of the main text. The tip was $Z_{\rm tip} = 10\ \mathrm{\upmu m}$ above the sample surface.}
    \label{fig:TypicalSpectra}
\end{figure}

\begin{figure}
    \centering
    \includegraphics[width=0.95\linewidth]{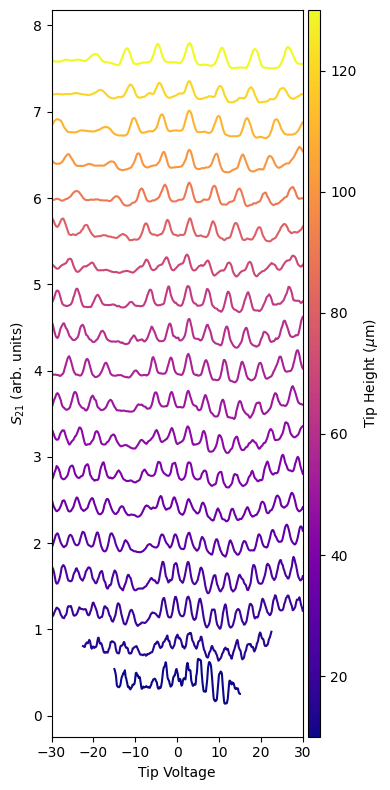}
    \caption{Keeping the tip above a Sisyphus defect and increasing the tip height reduces the electric field experienced by the Sisyphus defect, causing the spectral peaks to spread out. The data shows tip height variation from $Z_{\rm tip} = 10\ \mathrm{\upmu m}$ to $ 130\ \mathrm{\upmu m}$. The spectra were taken on the Sisyphus defect shown in Fig.~1 of the main text and are separated along the vertical axis for clarity. }
    \label{fig:SpectraHeight}
\end{figure}

\subsection{Power dependence}

\begin{figure*}
    \centering
    \includegraphics[width=1.0\linewidth]{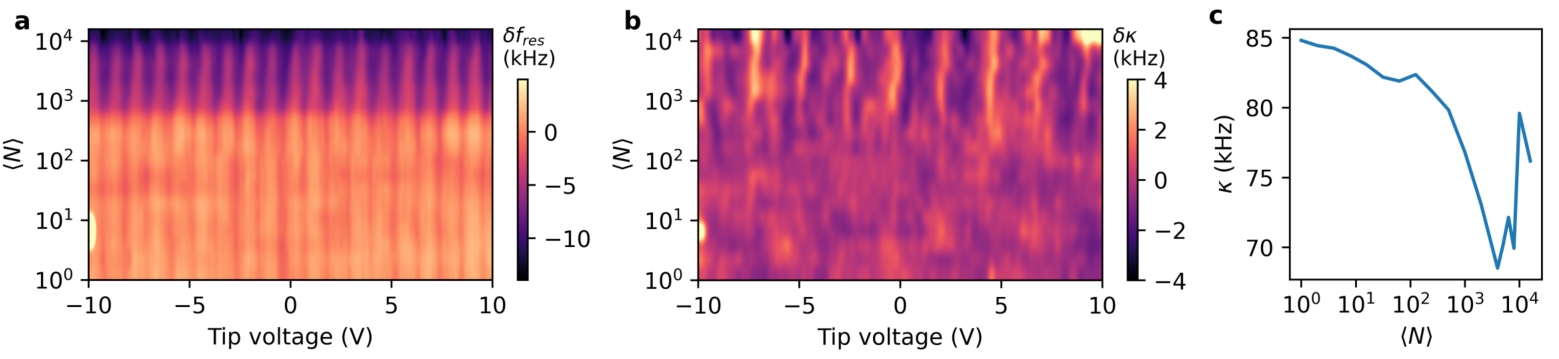}
    \caption{Power dependence of a Sisyphus defect. (a) same data as in the main manuscript showing the change in resonator frequency. (b) The complementary change in resonator loss rate due to the defect. The voltage-independent average resonator loss rate has been subtracted to flatten the data and remove the power-dependent average change in loss rate due to TLS saturation. (c) The calculated average loss rate $\kappa$ across all applied tip voltages, the background which has been subtracted from (b).}
    \label{fig:unsaturable}
\end{figure*}

\autoref{fig:unsaturable} shows the power dependence of the Sisyphus defects when resonator C's average photon number $\langle N\rangle$ is varied. \autoref{fig:unsaturable}a is the same data as in Fig.~2b of the main text for easy comparison, and \autoref{fig:unsaturable}b shows the periodic changes in resonator loss rate ($\delta\kappa$) extracted from the same dataset. Both panels had voltage-independent resonator loss at each photon number subtracted to accentuate the periodic features. The periodic frequency shifts persist unchanged up to $\langle N\rangle \sim 10^4$. The background contrast changes near $\langle N\rangle\sim 10^3$ because the resonator becomes non-linear when driven strongly. 

Increasing the number of photons in the resonator saturates the TLS defects coupled to it, reducing its loss rate, as shown in \autoref{fig:unsaturable}c. Above $\langle N\rangle\sim 10^4$, the resonator starts to become non-linear, and the loss rate ticks up, likely due to unreliable fitting.
Lower loss rates (higher Q) make the resonator more sensitive to small variations. Consistently, in \autoref{fig:unsaturable}b, periodic features are mainly seen at large $\langle N\rangle$. At lower photon occupancies, the loss due to the Sisyphus defect is washed out by TLS losses.

The doubling of the periodicity in \autoref{fig:unsaturable}a compared to \autoref{fig:unsaturable}b is likely because there are multiple defects close to each other, as shown in Figs. 2e and 4 of the main text. These defects likely lie in different regimes of the phase space shown in Fig.~3 of the main text, making some of them contribute more towards shifting the frequency of the resonator, while others contribute more to increasing the loss rate.

\section{Theoretical model}
 The observed physics can be described using a model consisting of a resonator that is capacitively coupled to a small island, which supports discrete charge states, that is tunnel-coupled to a continuum environment. The total Hamiltonian for this model can be written in the form
 \begin{equation}
    \hat{H} = \hat{H}_{c} + \hat{H}_R + \hat{H}_E + \hat{H}_I,
\end{equation}
where $\hat{H}_c$ is the charging Hamiltonian for the island (including an interaction term with the resonator), $\hat{H}_R$ is the free resonator Hamiltonian, $\hat{H}_E$ is the Hamiltonian describing the continuum environment, and $\hat{H}_I$ describes interactions between the island and electronic environment not including the resonator.
The resonator Hamiltonian can be written as
\begin{equation}
    \hat{H}_R = \hbar\omega_0\hat{a}^\dagger\hat{a},
\end{equation}
where $\hat{a}$ ($\hat{a}^\dagger$) are annihilation (creation) operators associated with the resonator,
and $\omega_0=1/\sqrt{LC}$ is the (angular) resonator frequency. 
The charging Hamiltonian of the island is given by \cite{Persson_2010}
\begin{equation}
\hat{H}_c = E_\mathrm c (\hat{n} - \hat{n}_\mathrm{g})^2,
\end{equation}
where $\hat{n}$ is the number operator for the island states, $\hat{n}_\mathrm{g}$ is the charge induced by an external voltage (including an interaction term containing the resonator microwave voltage operator), and $E_\mathrm{c} = e^2/2C_\Sigma$ is the charging energy with $C_\Sigma = C_\mathrm{j} + C_\mathrm{tip} + C_\mathrm{r}$ the total capacitance of the island. Here $\hat{n}_\mathrm{g} = n_\mathrm{g}^0 + n_\mathrm{g}^\mathrm{tip} + \hat{n}_\mathrm{g}^\mathrm{r}$ includes charge contributions due to external effects: $n_\mathrm{g}^0$ is the equilibrium charge, $n_\mathrm{g}^\mathrm{tip} = C_{\mathrm{tip}} V_{\mathrm{tip}}/e$ is the charge induced by the scanning probe tip electric field, and a contribution from the resonator interaction,
\begin{equation}
\hat{n}_\mathrm{g}^r = \frac{C_\mathrm{r} \hat{V}_\mathrm{rf}}{e}.
\end{equation}
The voltage operator can be written as 
\begin{equation}
    \hat{V}_\mathrm{rf} = V_\mathrm{rf} \hat{p}
\end{equation}
where $\hat{p} = \frac{i}{\sqrt{2}}(\hat{a}^\dagger - \hat{a})$ is a dimensionless momentum operator of the resonator, and $V_{\mathrm{rf}} = \omega_0 \sqrt{\hbar Z/2}$. Introducing the dimensionless coupling $\delta n_{\mathrm{g}} = C_\mathrm{r} V_\mathrm{rf}/e$ we have \cite{blais_2021}
\begin{equation}
\hat{n}_\mathrm{g}^r = \delta n_{\mathrm{g}} \hat{p}.
\end{equation}

In this work we consider an idealised model of the electronic bath environment as a non-interacting environment
\begin{equation}
    \hat{H}_E = \sum_k \epsilon_k \hat{c}^\dagger_k \hat{c}_k,
\end{equation}
where the energies $\epsilon_k$ of each environment mode can be obtained by discretising the bath density of states, and $\hat{c}_k$ ($\hat{c}^\dagger_k$) are the annihilation (creation) operators for an electron in the $k$-th bath mode. Finally, we consider an interaction Hamiltonian of the form \cite{PhysRevB.63.125315}
\begin{equation}
    \hat{H}_I = \left(\sum_{kl} V_{kl} \hat{c}_k^\dagger\hat{d}_l + H.c.\right),
\end{equation}
where $V_{kl}$ is the coupling between environment orbital $k$ and island orbital $l$, that captures electron transfer between the environment and the island. The sum over $l$ goes over all electronic states in the grain.

In what follows, we consider two separate approximate strategies for modelling the impact of the resonator--island coupling on the resonator dynamics:
\begin{enumerate}
    \setlength\itemsep{0pt}
    \item an Ehrenfest-type mean-field approach, which enables time-domain simulations of the coupled resonator--island dynamics and can account for resonator amplitude-dependent effects as well as contributions from multiple charge states,
    \item a linear-response treatment, which provides analytic expressions for the induced frequency shift and dissipation in the small resonator amplitude regime and reproduces the Ehrenfest results when the dot is restricted to the two most significant levels for the charge tunnelling processes (see Fig.~1c of the main text) and the resonator amplitude is small. 
\end{enumerate}
In addition, within both approaches,  the island dynamics will be treated within the Orthodox theory of single electron tunnelling \cite{AverinLikharev1991,752518}. This approach assumes that:
\begin{itemize}
    \setlength\itemsep{0pt}
    \item the system is in the incoherent, single electron tunnelling regime,
    \item the environment has a continuous, flat density of states ($\rho_{\mathrm{env}}$) with energy independent tunnelling rates (i.e. $V_{kl}=V$),
    \item the system-environment coupling is weak in the sense that $\rho_{\mathrm{env}} \rho_{\mathrm{dot}} V^2 \ll 1$, so that it can be treated at second order in perturbation theory; in terms of the tunnel resistance $R_\mathrm{t} = \frac{\hbar}{2\pi e^2}\frac{1}{\rho_{\mathrm{env}} \rho_{\mathrm{dot}} V^2}$ and the quantum of resistance $R_\mathrm{K} = \frac{h}{e^2}$, this condition of weak coupling is equivalent to $R_\mathrm t \gg R_\mathrm{K}$,
    \item the environment is Markovian.
\end{itemize}

A natural starting point for both the mean-field and linear response treatments is to use the Heisenberg equations of motion for the resonator variables
\begin{equation}
\begin{split}
    \dot{\hat{p}}(t) &= - \omega_0 \hat{q}(t), \\
    \dot{\hat{q}}(t) &=  \tilde{\omega}_0\hat{p}(t) - \frac{2 E_\mathrm c \delta n_{\mathrm{g}}}{\hbar}\left(\hat{n}(t) - (n_\mathrm{g}^0 + n_\mathrm{g}^\mathrm{tip})\right),
\end{split}
\end{equation}
where we have defined $\tilde{\omega}_0 = \left(\omega_0 - \frac{2E_\mathrm c \delta n_{\mathrm{g}}^2}{\hbar}\right)$ which accounts for a weak, static renormalisation of the resonator frequency due to coupling to the island, and $\hat{q}=(\hat{a}^\dagger+\hat{a})/\sqrt{2}$ is the resonator variable conjugate to $\hat{p}$.

\subsection{Ehrenfest Simulations}
In order to arrive at the Ehrenfest mean-field treatment for the coupled resonator--island dynamics, we now consider the evolution of the expectation values of the resonator variables
\begin{equation}
\begin{aligned}
\left\langle\dot{\hat{p}}(t)\right\rangle &= -\omega_0\!\left\langle \hat{q}(t)\right\rangle \\
\left\langle\dot{\hat{q}}(t)\right\rangle &= \tilde{\omega}_0\!\left\langle \hat{p}(t)\right\rangle -\frac{2 E_\mathrm{c} \delta n_{\mathrm{g}}}{\hbar}\left(\left\langle \hat{n}(t)\right\rangle\! -\! (n_\mathrm{g}^0 + n_\mathrm{g}^\mathrm{tip})\right)\!.
\end{aligned}\label{eq:odes}
\end{equation}
In order to arrive at a closed system of equations of motion we neglect correlations between island and resonator variables (e.g. $\left\langle \hat{n} \hat{p}\right\rangle \approx \left\langle \hat{n} \right\rangle \left\langle \hat{p}\right\rangle$). In doing so, we arrive at a set of classical equations of motion 
\begin{equation}
\begin{aligned}
\dot{p}(t) &= -\omega_0 q(t) \\
\dot{q}(t) &= \tilde{\omega}_0 p(t) -\frac{2 E_\mathrm{c} \delta n_{\mathrm{g}}}{\hbar}\left({N}(t) - (n_\mathrm{g}^0 + n_\mathrm{g}^\mathrm{tip})\right)\!.
\end{aligned}\label{eq:odes2}
\end{equation}
where $p(t)\equiv\left\langle \hat{p}(t)\right\rangle$ and $q(t)\equiv\left\langle \hat{q}(t)\right\rangle$ are classical variables representing the resonator state and $N(t)\equiv \left\langle \hat{n}(t)\right\rangle $ is the mean island charge. Now $N(t)$ can be obtained from the populations of charge state $n$, $P_n(t)$, as
\begin{equation}
\label{eq:nt}
N(t) = \sum_{n} n P_n(t).
\end{equation}
Within the Orthodox theory treatment we can describe the classical population of charge state $n$ by the rate equation \cite{AverinLikharev1991,752518}
\begin{equation}
\label{eq:pn}
\dot{P}_n = \sum_{n^\prime\neq n} \Gamma_{n^\prime\rightarrow n}(p) P_{n^\prime} - \sum_{n^\prime\neq n}\Gamma_{n\rightarrow n^\prime}(p) P_n, 
\end{equation} 
where $\Gamma_{n^\prime\rightarrow n}$ is the transition rate from charge states $n^\prime$ to state $n$ and depends on the instantaneous resonator field.  Similar expressions are found when considering coupling of single electron tunnelling to mechanical motion in suspended carbon nanotube devices \cite{PhysRevLett.115.206802, PhysRevResearch.4.043168}. Within the Orthodox theory treatment, the transition rate from charge state $n^\prime$ to state $n$ is \cite{752518, Persson_2010} 
\begin{equation}
\Gamma_{n^\prime\rightarrow n}(p) = \frac{1}{h} \frac{\Delta E_{n'\rightarrow n}(p)}{1 - \exp\left(\frac{\Delta E_{n'\rightarrow n}(p)}{k_\mathrm{B} T}\right)} \frac{R_\mathrm{K}}{R_\mathrm{t}}, 
\end{equation}
where $\Delta E_{n'\rightarrow n}(p) = E_{n'}(p) - E_{n}(p)$ is the energy difference between the two states and each energy is given by
\begin{equation}
E_n(p) = E_\mathrm{c}\left[n-\left(n_\mathrm{g}^0 + n_\mathrm{g}^\mathrm{tip}+\frac{C_\mathrm{r} V_\mathrm{rf}}{e}p\right)\right]^2. 
\end{equation}
We numerically solve the coupled system of ODEs (Eqs.~\ref{eq:odes2} and \ref{eq:pn}) for the time-dependent resonator expectation values and defect state populations, from which we extract the resonator frequency and decay rates.

\subsection{Linear Response Treatment}
Assuming that the island--resonator coupling is weak ($\delta n_{\mathrm{g}}\ll1$), we can treat the interaction as a small perturbation, and apply linear response theory \cite{Kubo_1966, Riseborough_1985}.  
From the Heisenberg equations of motion, and noting that to first order in $\delta n_{\mathrm{g}}$ we have $\tilde{\omega}_0 \approx \omega_0$, for the resonator $\hat{q}$ and $\hat{p}$ variables we obtain 
\begin{equation}
    \ddot{\hat{p}} + \omega_{\mathrm{0}}^2 \hat{p} = \frac{2 \omega_0 E_\mathrm c \delta n_{\mathrm{g}}}{\hbar} \left(\hat{n}(t) - (n_{\mathrm{g}}^0+n_{\mathrm{g}}^{\mathrm{tip}})\right).
\end{equation}
 Separating the charge into time-dependent and time-independent contributions $\hat{n}(t) = \hat{n}_0 + \delta \hat{n}(t)$, and noting that any time-independent contribution contributes to a static offset in $\hat{p}$, and can therefore be absorbed by redefining $\hat{p}$, we find that
\begin{equation}
    \ddot{\hat{p}} + \omega_{0}^2 \hat{p} = \frac{2 \omega_0 E_\mathrm c \delta n_{\mathrm{g}}}{\hbar} \delta \hat{n}. \label{eq:linear_response_ode}
\end{equation}

We may write the induced charge fluctuation $\delta \langle \hat{n}(t) \rangle$ due to resonator-island coupling term as \cite{Kubo_1966, Riseborough_1985}
\begin{equation}
    \delta \langle \hat{n}(t) \rangle = -2 E_\mathrm c \delta n_{\mathrm{g}} \int_{-\infty}^t  \chi_{nn}(t-\tau) \langle \hat{p}(\tau) \rangle \mathrm{d}\tau, \label{eq:charge_fluc}
\end{equation}
where 
\begin{equation}
    \chi_{nn}(t) = \frac{i}{\hbar} \Theta(t) \left\langle \left[\hat{n}(t), \hat{n}(0)\right] \right\rangle
\end{equation}
is the charge susceptibility of the island.  Inserting Eq.~\ref{eq:charge_fluc} into Eq.~\ref{eq:linear_response_ode} and converting to Fourier space we find
\begin{equation}
    \left[-\omega^2+\omega_0^2 + \frac{ (2 E_\mathrm c\delta n_{\mathrm{g}})^2 \omega_0}{\hbar}\chi_{nn}(\omega)\right] \left\langle\hat{p}(\omega)\right\rangle = 0.
\end{equation}
For an arbitrary perturbation $\left\langle\hat{p}(\omega)\right\rangle$ we can find a solution to this equation by requiring that the term in square brackets is zero, which by expanding around the bare resonator frequency and assuming that the resonator coupling weakly perturbs the system (allowing us to only retain the lowest order terms in the perturbation) gives the frequency shift and excess dissipation as 
\begin{align}
    \delta \omega &=  \frac{1}{2\hbar}(2 E_\mathrm c\delta n_{\mathrm{g}})^2 \mathrm{Re}\left[\chi_{nn}(\omega_0)\right],\\
    \delta \kappa &=  -\frac{1}{\hbar}(2 E_\mathrm c\delta n_{\mathrm{g}})^2 \mathrm{Im}\left[\chi_{nn}(\omega_0)\right].
\end{align}

In order to evaluate the charge susceptibility, we assume that the island population dynamics can be described using Orthodox theory applied to the two most relevant energy levels (namely the two charge states with $\langle \hat{n} \rangle$ closest in value to $n_{\mathrm{g}}^0+n_{\mathrm{g}}^{\mathrm{tip}}$, see Fig.~1c of the main text). Within these approximations the charge population dynamics of the island simplifies to
\begin{equation}
\begin{split}
    \dot{P}_1 &= \Gamma^-(\left\langle \hat{p}\right\rangle) P_0 - \Gamma^+(\left\langle \hat{p}\right\rangle) P_1 \\
    \dot{P}_0 &= \Gamma^+(\left\langle \hat{p}\right\rangle) P_1 - \Gamma^-(\left\langle \hat{p}\right\rangle) P_0,
\end{split}
\end{equation}
where we have introduced $P_{0}$ and $P_{1}$ as the population on charge state $n$ and $n+1$, respectively, $\Gamma^+(\left\langle \hat{p}\right\rangle) = \Gamma_{n+1\rightarrow n}(\left\langle \hat{p}\right\rangle)$ and 
$\Gamma^-(\left\langle \hat{p}\right\rangle) = \Gamma_{n\rightarrow n+1}(\left\langle \hat{p}\right\rangle)$, 
and where the energy difference is
\begin{equation}
    \Delta E_{n+1, n}(\left\langle \hat{p}\right\rangle) = E_\mathrm c\left[1 + 2(n-(n_{\mathrm{g}}^0 + n_{\mathrm{g}}^\mathrm{tip} ))\right] - 2 E_\mathrm c \delta n_{\mathrm{g}}  (\left\langle \hat{p}\right\rangle).
\end{equation} 

Using that $P_0+P_1=1$ we arrive at
\begin{equation}
\dot{P}_1 = \Gamma^- - \Gamma P_1,
\end{equation}
where $\Gamma = \Gamma^+ + \Gamma^-$
and at equilibrium $P_1 = \frac{\Gamma^-}{\Gamma}$. Next we assume that the resonator coupling perturbs the dot dynamics as 
\begin{equation}
    P_1(t) = P^{0}+\delta P(t),
\end{equation}
where $P^0 = \Gamma^-(\Delta E_0)/\Gamma(\Delta E_0)$ is the equilibrium population in the absence of any perturbation.  From this we see that $\delta \langle \hat{n}(t) \rangle = \delta P(t)$. We then write the energy difference between the two states in terms of a time-independent part  ($\Delta E_0$) and the time-dependent perturbuation ($\delta E(\left\langle \hat{p}\right\rangle) = -2E_\mathrm c \delta n_{\mathrm{g}}\left\langle \hat{p}\right\rangle $) as $\Delta E(\left\langle \hat{p}\right\rangle) = \Delta E_0 + \delta E(\left\langle \hat{p}\right\rangle)$.  Now expanding the rates to first order in the perturbation
\begin{equation}
    \Gamma^\pm \approx \Gamma^{\pm}(\Delta E_0) + \frac{\mathrm{d}\Gamma^{\pm}}{\mathrm{d} E}\Bigg|_{\Delta E_0}\delta E
\end{equation}
and inserting into the Master equation for the population dynamics we obtain
\begin{equation}
    \dot{\delta P} = - \Gamma(\Delta E_0) \delta P - 2\delta n_{\mathrm{g}} E_\mathrm c \left\langle \hat{p}(t)\right\rangle \Gamma(\Delta E_0) \frac{\mathrm{d}P^0}{\mathrm{d}  E}\Bigg|_{\Delta E_0},
\end{equation}
where we have discarded all terms that are second order in the perturbation and where we used the fact that 
\begin{equation}
    \frac{\mathrm{d}P^{0}}{\mathrm{d}  E}\Bigg|_{\Delta E_0} = \frac{1}{\Gamma(\Delta E_0)}\left(\frac{\mathrm{d}\Gamma^-}{ \mathrm{d} E}\Bigg|_{\Delta E_0}-P^0\frac{\mathrm{d}\Gamma}{\mathrm{d} E}\Bigg|_{\Delta E_0}\right).
\end{equation}
We transform this to the frequency domain using the Fourier transform, and solve it to obtain the following relation for the charge fluctuation,
\begin{equation}
    \delta N(\omega) =-2 \delta n_{\mathrm{g}} E_\mathrm c \frac{\Gamma(\Delta E_0)}{\Gamma(\Delta E_0) - i\omega}  \frac{\mathrm{d}P^{0}}{\mathrm{d} E}\Bigg|_{\Delta E_0} \left\langle \hat{p}(\omega)\right\rangle,
\end{equation}
giving
\begin{equation}
    \chi_{nn}(\omega) = \frac{\Gamma(\Delta E_0)}{\Gamma(\Delta E_0) - i\omega}  \frac{\mathrm{d}P^{0}}{\mathrm{d} E}\Bigg|_{\Delta E_0}. \label{eq:suscept}
\end{equation}

From this we obtain the following analytic expressions for the frequency shift and excess dissipation of the resonator due to the coupling to the island charge states
\begin{align}
    \delta \omega &=  \frac{1}{\hbar}\left(\frac{2 E_\mathrm c V_\mathrm{rf} C_\mathrm{r}}{e} \right)^2 \frac{\Gamma(\Delta E_0)^2}{\Gamma(\Delta E_0)^2 + \omega_0^2} \frac{\mathrm{d}P^{0}}{\mathrm{d} E}\Bigg|_{\Delta E_0}\\
    \delta \kappa &= -\frac{2}{\hbar}\left(\frac{2 E_\mathrm c V_\mathrm{rf} C_\mathrm{r}}{e} \right)^2  \frac{ \omega_0 \Gamma(\Delta E_0)}{\Gamma(\Delta E_0)^2 + \omega_0^2} \frac{\mathrm{d}P^{0}}{\mathrm{d} E} \Bigg|_{\Delta E_0},
\end{align}
where 
\begin{equation}
\Gamma(\Delta E_0) = \frac{\Delta E_0}{h}\frac{R_\mathrm{K}}{R_\mathrm t} \coth
\left(\frac{\Delta E_0}{2k_\mathrm{B} T}\right)\label{eq:gamma}
\end{equation}
and
\begin{equation}
\frac{\mathrm{d}P^{0}}{\mathrm{d} E} \Bigg|_{\Delta E_0} = -\frac{1}{4 k_\mathrm{B} T} \mathrm{sech}^2\left(\frac{\Delta E_0}{2k_\mathrm{B} T}\right). \label{eq:thermal_contribution}
\end{equation}

Finally, we can write the frequency shifts in terms of the resonator frequency and impedance, giving:
\begin{align}
    \delta \omega &=\! -  \frac{Z}{2 k_\mathrm{B} T}\!\!\left(\!\frac{E_\mathrm c C_\mathrm{r}}{e}\!\right)^2 \!\!\!\!\omega_0^2\frac{\Gamma(\Delta E_0)^2}{\Gamma(\Delta E_0)^2 + \omega_0^2} \mathrm{sech}^2\!\left(\!\frac{\Delta E_0}{2k_\mathrm{B} T}\!\right)\label{eq:dw}\\
    \delta \kappa &= \frac{Z}{k_\mathrm{B} T}\!\!\left(\!\frac{E_\mathrm c C_\mathrm{r}}{e}\!\right)^2 \!\!\!\!\omega_0^2 \frac{\omega_0 \Gamma(\Delta E_0)}{\Gamma(\Delta E_0)^2 + \omega_0^2} \mathrm{sech}^2\!\left(\!\frac{\Delta E_0}{2k_\mathrm{B} T}\!\right)\label{eq:dk}.
\end{align}

\section{Theoretical discussion}
Near a charge degeneracy point (i.e. $\{n_\mathrm{g}^0 + n_\mathrm{g}^\mathrm{tip}\} \approx 0.5$), the RF voltage from the resonator alters which of the two charge states is the lowest energy state, following which an electron tunnelling event can occur. This tunnelling event dissipates energy, which can be modelled as an effective Sisyphus resistance \cite{Persson_2010}. A change of the defect charge state results in a back-action on the resonator, which, to the lowest order, shifts the resonator frequency that we measure.  

An important distinction from Ref. \cite{Persson_2010} is that here we measure the resonator response instead of directly monitoring the island charge state dynamics. Thus, in our case, the back-action on the resonator from the Sisyphus defect determines the measured response. As we probe the resonator rather than the defect itself, we observe the periodic response to be power independent (Fig.~2b of the main text). This observation is consistent with simulations (and linear response expressions Eqs.~\ref{eq:dw} and ~\ref{eq:dk}), which predict the response to be independent of driving power at low and moderate photon numbers (up to $\sim 10^6$). The simulations also predict that driving harder should broaden the peaks due to effects arising beyond linear response in the resonator-island coupling, but due to the hardware configuration that was not possible to explore experimentally.

From Eqs. (\ref{eq:gamma}),  (\ref{eq:dw}) and (\ref{eq:dk}) we see that the response scales differently with resonator frequency $\omega_0$ depending on the timescales of electron tunnelling dynamics $hR_\mathrm t/\Delta E_0 R_\mathrm{K}$. For frequencies exceeding this timescale Eqs.  (\ref{eq:dw}) and (\ref{eq:dk}) asymptotically take on a constant and $\delta \kappa\propto \omega_0$ dependence respectively, meaning, as long as the physics of single electron tunnelling holds (for sufficiently large $R_\mathrm t$) this mechanism is relevant for all device frequencies. Relevant to this is also the size distribution of possible grains present, which may be different for different materials. We expect there to be both a lower and an upper limit to grain sizes, which means there will be constraints on the ranges of possible $R_\mathrm t$ in a device.
From Eq. (\ref{eq:dk}) we also derive the $\propto 1/\omega_0$ dependence in Fig.~3 for the $R_\mathrm t$ that maximally contribute to the dissipation.

\subsection{Peak broadening}
In this section we discuss the broadening of the Sisyphus defect absorption peaks. We start by considering the relative broadening with respect to the gate oscillation period. We can express this as the voltage broadening
\begin{equation}
    \Delta V_\mathrm{g, b} = \frac{\Delta E_\mathrm{b}}{\alpha e},
\end{equation}
where $\Delta E_\mathrm{b}$ is the broadening in energy units and $\alpha\equiv C_\mathrm{g}/C_\Sigma$ is the capacitive lever arm of the gate capacitance to the total capacitance of the island. Similarly, the voltage period is
\begin{equation}
    \Delta V_\mathrm{g, p} =\frac{e}{\alpha C_\Sigma}.
\end{equation}
Consequently, using $E_\mathrm{c} = e^2/2C_\Sigma$, the relative broadening becomes 
\begin{equation}
    \frac{\Delta V_\mathrm{g, b}}{\Delta V_\mathrm{g, p}} = \frac{\Delta E_\mathrm{b}}{2E_\mathrm{c}}.
\end{equation}
From this result we see that the probability that a Sisyphus defect is `active' and interacts with the quantum circuit does not depend on the lever arm $\alpha$ or the gate capacitance $C_\mathrm{g}$ (assuming $C_\mathrm{g}\ll C_\Sigma$). This means that this ratio is independent of tip position, and also holds in the absence of a tip, irrespective of the underlying physical mechanism for the peak broadening. 

We next discuss different factors that can contribute to the absolute broadening.\\

{\bf{Thermal and lifetime broadening.}} 
From the linear response treatment, the frequency shift (Eq.~\ref{eq:dw}) and dissipation (Eq.~\ref{eq:dk}) are proportional to the real and imaginary parts, respectively, of Eq.~\ref{eq:suscept}.  This expression contains two important terms. The first term contains all temporal effects, that is it accounts for the finite tunnelling time on the dot, the second contains the thermal broadening and largely controls the width of the response function. Considering only the thermal broadening we find that the peak will have the shape (see Eq. \ref{eq:dk})
\begin{equation}
    \mathrm{sech}^2\!\left(\!\frac{\Delta E_0}{2k_\mathrm{B} T}\!\right).
\end{equation}
This is analogous to the well-known Beenakker peak broadening result for single electron transistors in the weak tunnelling limit \cite{Beenakker_1991}. It gives a full width half maximum in $n_\mathrm{g}$ of $\updelta n_\mathrm{g,FWHM}=3.53 k_\mathrm{B}T/(2E_\mathrm{c})$. For the defect in the main text Fig.~2f with $C_\mathrm{j}=0.35$ fF at $T=200$ mK, this gives $\updelta n_\mathrm{g,FWHM}\approx 0.25$,

which is comparable to what we see experimentally. Similarly we estimate for $C_\mathrm{j}=0.1$ fF  and  $T=100$ mK that $\updelta n_\mathrm{g,FWHM}\approx 0.04$, the conditions of the data in Fig.~2d. This is about 3 times smaller than the experimentally observed width, suggesting that the width of this defect could be limited by a different mechanism. Indeed, we also did not observe any appreciable change in the peak width of this defect up to $T=1.1$ K, as shown in \autoref{fig:temp_defect_a}. 

We note that the measured fridge temperature may underestimate the temperature of the local Sisyphus defects. Environmental degrees of freedom in solid state circuits tend to thermalise to temperatures in excess of the base temperature of the fridge, typically at 40--50 mK. This means that even if limited by thermal broadening, the probability of a defect remaining `active' remains high, even if a defect were to only be limited by thermal broadening at 10 mK.

\begin{figure}[ht!]
\centering
\includegraphics[width=0.45\textwidth]{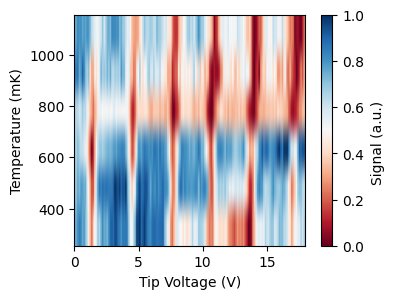}
\caption{Temperature dependence of the peaks from the defect in Fig.~1 of the main text. The tip was kept at a fixed position near the defect while the temperature of the refrigerator was swept. To remove effects of resonator parameters changing as a function of temperature, the signal has been normalised. }
\label{fig:temp_defect_a}
\end{figure}

{\bf{Broadening due to high-frequency environmental noise.}} Consider environmental charge noise producing local potential fluctuations $\updelta\phi_\mathrm{env}$ which shifts the island energy by $\updelta E = e\updelta\phi_\mathrm{island}$. The magnitude of this noise can in principle be arbitrarily large; depending on the local charge environment individual fluctuations could cause variations in the induced island charge corresponding to a significant fraction of $1e$, as we also observe through occasional slow charge jumps in our data. In the presence of a SGM tip the island potential is given by the capacitive divider between the environment, the tip, and the rest of the island capacitances to surrounding circuitry. We introduce $C_{\Sigma'}$ such that $C_\Sigma = C_{\Sigma'} + C_\mathrm{tip}(r)$ to explicitly show the influence of the tip-island geometry such that
\begin{equation}
    \updelta\phi_\mathrm{island} = \frac{C_\mathrm{env}\phi_\mathrm{env}+C_\mathrm{tip}(r)\phi_\mathrm{t}}{C_{\Sigma'}+C_\mathrm{tip}(r)},
\end{equation}
where $C_\mathrm{env}$ is the effective capacitance between the island and its noisy local environment.
If we assume a static tip potential we can write down the fluctuating contribution to the dot energy
\begin{equation}
    \updelta E(r) = e\updelta \phi_\mathrm{env}\frac{C_\mathrm{env}}{C_{\Sigma'}+C_\mathrm{tip}(r)}.
\end{equation}
We can express this as a peak broadening in voltage
\begin{equation}
    \Delta V_\mathrm{g,env}(r) = \frac{\updelta E(r)}{\alpha(r)e} = \updelta \phi_\mathrm{env} \frac{C_\mathrm{env}}{C_\mathrm{tip}(r)},
\end{equation}
by considering the capacitive lever arm to the tip.
This result can easily be understood from the tip acting as a capacitive divider, and when close to the island it absorbs a significant part of the environmental fluctuations. Expressed in terms of relative broadening we find
\begin{equation}
    \frac{\Delta V_\mathrm{g,env}(z)}{\Delta V_{g, p}} = \updelta \phi_\mathrm{env} \frac{C_\mathrm{env}}{e},
\end{equation}
which is also independent of the tip location to first approximation.

\begin{figure}
\centering
\includegraphics[width=0.4\textwidth]{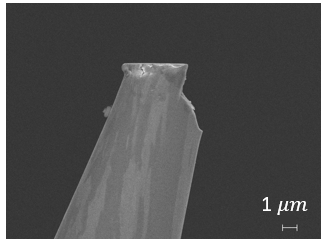}
\caption{Scanning electron micrograph of the tip taken after conducting the measurements in \autoref{fig:ctg}.}
\label{fig:tip}
\end{figure}

\begin{figure*}
\centering
\includegraphics[width=0.95\textwidth]{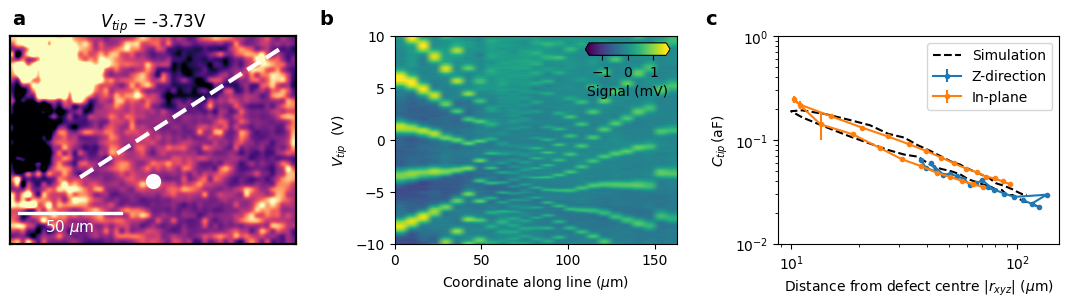}
\caption{ (a) Grid image taken prior to the data in panels (b) and (c). The dashed line shows the line along which the data in panel (b) was taken. The white dot indicates the location of the Z-direction data in panel (c). (b) Measured readout signal as a function of tip voltage taken at points along the line indicated in (a), passing through the centre of the rings where the defect is located. The tip was lifted $Z_{\rm tip} = 10\ \mathrm{\upmu m}$. The periodicity of the peaks in voltage are used to evaluate the tip--island capacitance $C_\mathrm{tip}$ in (c). (c) Extracted tip-grain capacitance as a function of distance from the defect (centre of rings). The dashed line shows the result of an electrostatic simulation (see text). Orange data are from moving the tip along the dashed line in (a). Blue data are data obtained when moving the tip in $Z$ at the location of the dot in panel (a). Error bars indicate the standard deviation of voltage between individual peaks detected, and the data points are the means.}
\label{fig:ctg}
\end{figure*}

\section{Island--tip capacitance \& modelling}
The periodic response measured in the $S_{21}(f_{\rm readout})$ spectrum as a function of applied tip voltage $V_{\mathrm{tip}}$ as shown in \autoref{fig:TypicalSpectra} is a measure of the tip-grain capacitance $C_\mathrm{tip}=e/\Delta V_{\mathrm{tip}}$. Hence, measuring how this periodicity varies as a function of tip location can be used to estimate the size of a Sisyphus defect if the tip size is known. We estimate the tip size from scanning electron microscopy (SEM) images taken after performing the measurements for the defect in Fig.~1 of the main text (\autoref{fig:tip}). With the tip placed $Z_{\rm tip} = 10\ \mathrm{\upmu m}$ directly above the defect in Fig.~1 of the main text, a charge state periodicity of $\Delta V_{\rm tip} = 0.7$ V is observed in a $S_{21}(f_{\rm readout})$ spectrum, implying $C_\mathrm{tip} = 2.3\times 10^{-19}$ F assuming the grain to be non-superconducting.

No smearing of the spectra was observed as the sample temperature was increased to 1 K, meaning the Coulomb blockade energy $E_\mathrm c = e^2/2C_{\Sigma}\gg k_\mathrm{B}T$, where $C_{\Sigma}$ is the total capacitance of the island and the coupling capacitance of the island to the ground plane, $C_\mathrm{j}$, is the dominant contributor to $C_{\Sigma}$ for all tip positions in our experiment.

\autoref{fig:ctg} shows the extracted tip--island capacitance $C_\mathrm{tip}$ as the tip was moved along a line traversing the defect location at a height of $Z_{\rm tip} = 10\ \mathrm{\upmu m}$, as well as the capacitance extracted from keeping the tip at a fixed point in the xy-plane and moving it in the z-direction. \autoref{fig:ctg}b shows the variation of the spectra as the tip was moved parallel to the sample plane, while \autoref{fig:SpectraHeight} shows the variation when the tip was moved perpendicular to the sample. \autoref{fig:ctg}c shows that the two measurements are in good agreement with each other, provided we take into account the full distance of the tip $r_{xyz} = (\Delta x^2 + \Delta y^2 + \Delta z^2)^{1/2}$ from the defect location. This observation implies that the Sisyphus defect was located near the sample surface, rather than being embedded within the substrate.

To estimate the size of the defect, an electrostatic simulation in COMSOL was implemented, modelling the tip as a $50\ \mathrm{\upmu m}$ radius wire terminated by a conical tip with a flat apex of radius $1.5\ \mathrm{\upmu m}$ and height $70\ \mathrm{\upmu m}$. This is an approximate implementation of the tip as observed in SEM (\autoref{fig:tip}). The defect was modelled as a disc with radius $r_\mathrm{g}$ located on the silicon substrate, next to a grounded metal electrode. $r_\mathrm{g} = 12$ nm is required to match the simulation with experimental data \autoref{fig:ctg}c). We note that this grain size of 12 nm is only an estimate, as the precise size and shape of the tip at the time of the measurements shown in \autoref{fig:ctg} is not possible to obtain accurately. 

We also note an asymmetry in the in-plane measurement in \autoref{fig:ctg}c, where the capacitance curve for the tip approaching the defect is slightly offset from the one for the tip moving away from it. We can reproduce the effect in our simulations by introducing large grounded metallic structures in the device and placing the defect closer to one of them. Due to the non-symmetric placement of the defect with respect to larger grounded features on the sample, the measured $C_\mathrm{tip}$ vs tip position becomes asymmetric as a result of the screening and wider capacitive network introduced by other metallic structures on the sample.

\subsection{Estimated Coulomb blockade properties of Nb grains}
Taking the dielectric constant of NbO $\varepsilon_\mathrm{r} = 30$, and an oxide thickness of $d=2$ nm we can estimate the junction capacitance and the charging energy of a grain of size $r=11$ nm, as per the X-ray results in the next section. Assuming the junction is formed by half of the grain we get $C_\mathrm{j} = 2\pi \varepsilon_0\varepsilon_\mathrm{r} r^2/d = 0.1$ fF, in agreement to the theoretical result yielding a good match to the experimental data for $0.1$ fF. The self-capacitance of such grain $C_\mathrm{self}=4 \pi \varepsilon_0r\approx 1.2$ aF is much smaller such that the total capacitance $C_\Sigma$ is determined by $C_\mathrm{j}$. Based on previous studies of Nb oxide tunnel barriers \cite{Im_2007, Im_2006}, for a grain and oxide barrier of this size, we also arrive at an order of magnitude estimate for the tunnel resistance $R_\mathrm{t}$ in the range of 100 k$\Omega$. However, the properties of Nb oxide can vary significantly, and has not been extensively studied in the context of tunnel junctions, and are likely to vary significantly even across the same device depending on local geometry and composition.

\section{Ex-situ materials science}
In this section, we outline the results of applying a range of different ex-situ characterisation techniques to the Nb on Si thin films in which we found multiple Sisyphus defects.

\subsection{X-ray scattering}
X-ray reflectometry (XRR) measurements were carried out on a sample that was cut from the wafer from on which resonators B and C were fabricated. The structure of the Nb thin film grown on a silicon substrate was determined using the Bruker D8 Discover Multipurpose X-ray Diffractometer at Royal Holloway using Cu K$\alpha$ radiation. The specular X-ray reflectivity was measured with a G\"obel mirror and a fixed slit in the incident beam path, and with a 0.225 mm receiving slit in front of the LynxEye XE-T detector in 0D mode. Standard powder diffraction measurements were performed in the Bragg--Brentano scattering geometry with variable slit and fixed sample illumination of 10 mm, and with the detector in 1D mode. Grazing-incidence X-ray diffraction measurements were performed with a G\"obel mirror and fixed slit in the incident path, and with a $0.3^\circ$ equatorial Soller and 14.325 mm receiving slit in front of the detector in 0D mode. Scans of detector angle were performed with a fixed incident angle of $0.5^\circ$. The results are presented in \autoref{fig:XRR}.

\begin{figure*}
\centering
\includegraphics[width=0.7\textwidth]{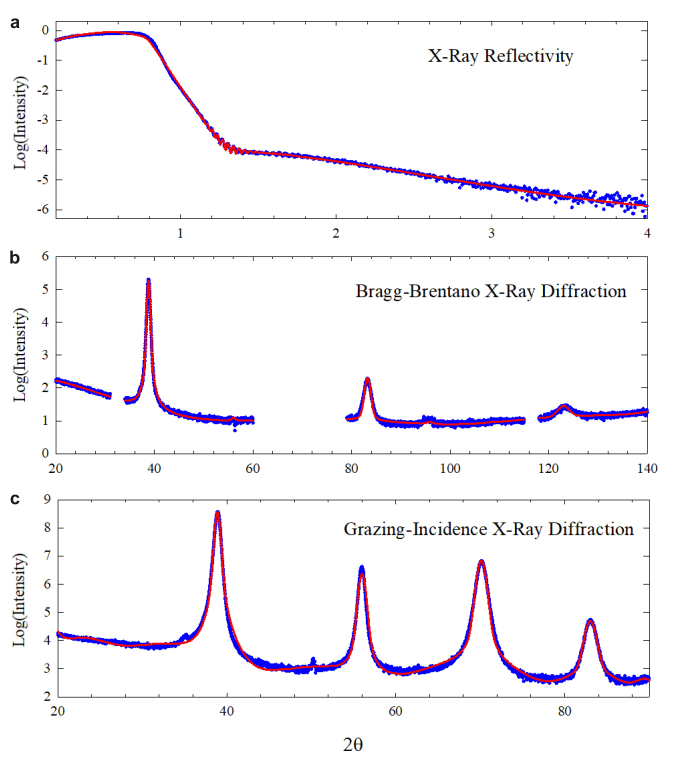}
\caption{(a) The X-ray reflectivity, (b) Bragg--Brentano X-ray diffraction, and (c) the grazing-incidence X-ray diffraction from a Nb film on a silicon substrate. The blue points show the experimental data, and the red lines show the fits. The regions dominated by scattering from the substrate are omitted for clarity.}
\label{fig:XRR}
\end{figure*}

Fits to the reflectivity data were performed using the Leptos software using the virtual diffractometer to calculate the resolution function on the basis of the X-ray optical beam path. It is necessary to include a surface oxide layer in addition to the Nb layer. There are two oscillation scales in the data. The rapid oscillations correspond to the Nb film thickness, $204(3)$ nm, and the broad feature gives a surface oxide thickness of $7(3)$ nm, with a large surface roughness of $3.5(7)$ nm. 
Rietveld refinement of the Bragg--Brentano X-ray diffraction data were performed using the Topas software, where the instrumental resolution was determined in a fundamental parameters approach, so that the details of the microstructure are obtained on an absolute scale. The fit shows that the polycrystalline Nb film has the Im$\bar{3}$m space group with a lattice parameter of 0.3282 nm, and it has preferred orientation. The crystallite size in the direction perpendicular to the surface is $20.94(15)$ nm. Regions of very strong scattering from the substrate are omitted from the figure for clarity.
The grazing-incidence X-ray diffraction focusses on the surface region, limiting scattering from the substrate to a few very weak sharp features. It shows the expected Bragg reflections from the Nb film. The absence of any further broad peaks suggests that the surface oxide layer is amorphous. Rietveld refinement was again performed using the Topas software. In this case, the instrumental resolution was calibrated using a corundum standard sample, and a crystallite size of $21.2(2)$ nm fits the data well across the whole scan, where each peak corresponds to a different direction relative to the surface normal. This is consistent with the result from Bragg--Brentano X-ray diffraction, and it suggests that the average crystallite size is 21 nm in all directions.

\subsection{AFM}
In \autoref{fig:afm} we show an atomic force microscopy (AFM) image taken atop the Nb surface of one of our devices. These data are in agreement with the X-ray data, showing surface variations of sub-100 nm size and a surface peak-to-peak roughness of $\sim 3$ nm, in good agreement with the X-ray data.

\begin{figure*}[ht!]
\centering
\includegraphics[width=0.9\textwidth]{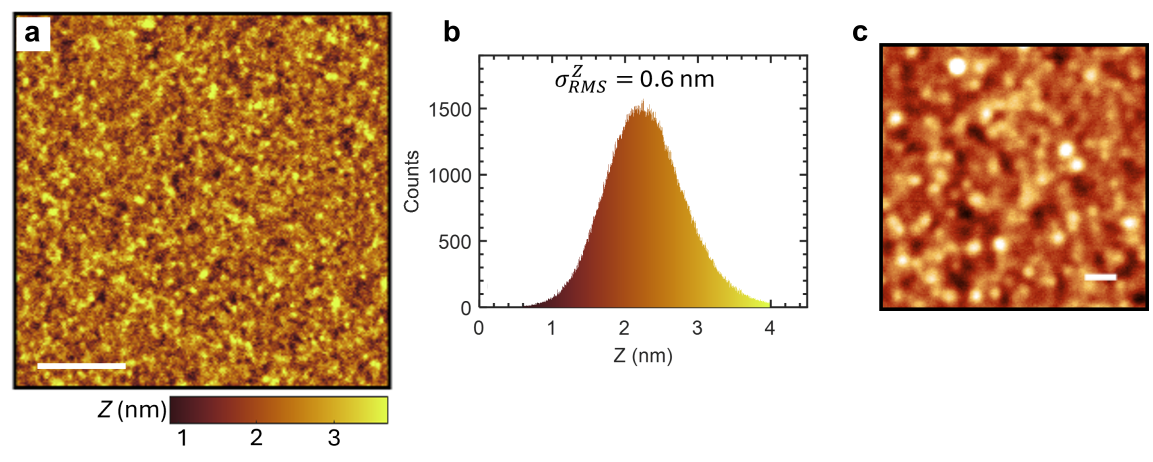}
\caption{Atomic force microscopy of the Nb surface. (a) AFM topography. The scale bar is 1 $\upmu$m. Data taken at room temperature. (b) Histogram of measured heights from the image in (a). (c) close-up image showing the AFM topography on the Nb surface. Scale bar is 50 nm.}
\label{fig:afm}
\end{figure*}

\subsection{SEM \& EDS}
Granularity near the edges of patterned features can also be seen in SEM. \autoref{fig:SEM} shows example images of our films taken using scanning electron microscopy (SEM) and energy dispersive spectroscopy (EDS). An image of a milled-out section of the device in \autoref{fig:SEM} a shows the sample cross-section where a slight overhang of the Nb film above the Si substrate is visible. The granularity shown along a thin-film edge in \autoref{fig:SEM}b is consistent with the typical grain size, and also some larger grains are visible near edges. EDS shows the expected materials composition of Nb and Nb oxide.  

\begin{figure*}
\centering
\includegraphics[width=0.75\textwidth]{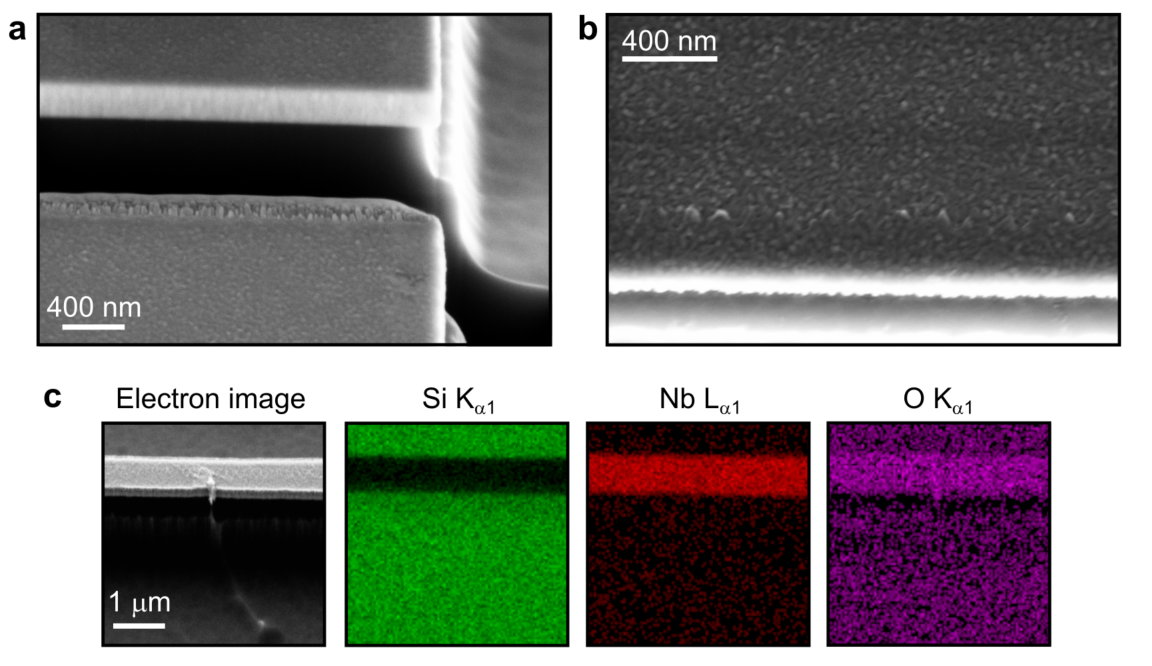}
\caption{SEM and EDS on the Nb on Si sample (device B and C). (a) SEM image of an in-situ milled cross-section showing the Nb film (200 nm) and trench profile ($1.2\ \mathrm{\upmu m}$ trench depth). (b) SEM image of lithographically defined edge (bottom bright contrast) and Nb surface showing grains in the Nb film, edge surface roughness of similar scale as Nb grains, and some additional deposits on the Nb film near the edge. (c) Energy dispersive spectroscopy (EDS) of an lithographically defined edge of the Nb film. Panels show the electron image, and the intensity of detected energy-specific X-ray signal corresponding to the Si $K_{\alpha 1}$, Nb $L_{\alpha 1}$, and O $K_{\alpha 1}$ peaks respectively. The image was taken at an angle of $45^\circ$ with respect to the sample surface.}
\label{fig:SEM}
\end{figure*}


\section{Fourier analysis and separation of distinct Sisyphus defects}
\label{sec:SIfourier}
We now describe in detail the method and data analysis for detecting Sisyphus defects using a Fourier transform technique, first applied to the dataset from Resonator C.

Because the periodic resonator response of a Sisyphus defect with its potential tuned by the applied tip voltage depends on the tip--defect capacitance, $\Delta V_\mathrm{tip} = q / C_\mathrm{tip}$, the periodicity depends on the distance $|\vec{r}|$ of the tip from the defect.
A Fourier transform of each $V_\mathrm{tip}$ sweep recorded in each position of the tip in a grid can therefore  isolate periodic components arising from independent tip--defect interactions, removing non-periodic backgrounds and revealing weak tip--defect interactions that are otherwise masked by stronger signals.
For convenience, we work with the reciprocal voltage frequency $\nu = 1/ \Delta V_\mathrm{tip}$ (units V$^{-1}$).
In the frequency domain we expect a periodic response from a Sisyphus defect (an observed ridge in the spectrum) following a $\nu \propto |\vec{r}|^{-\beta}$ relation relative to the ridge maximum (where the tip is closest to the defect), with constant $\beta$. This enables the separation of distinct defects, even if their responses overlap with signals from other defects at certain scan locations.
It is also possible to use this as a method for triangulating defect locations, even for defects lying outside of the scan area.

\begin{figure*}[htbp]
\centering
\includegraphics[width=0.99\textwidth]{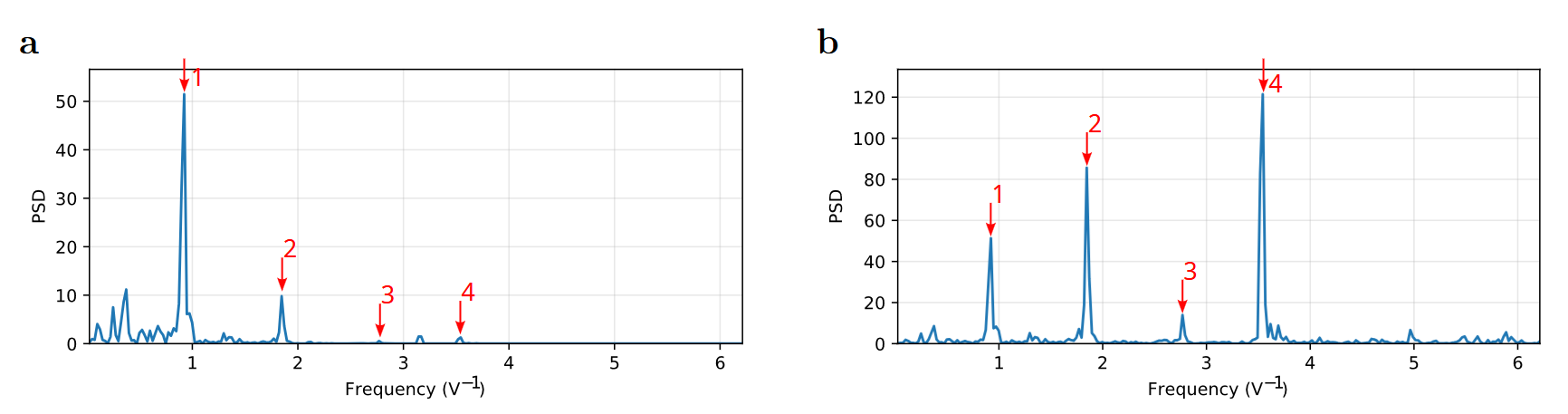}
\caption{
  (a) The raw PSD for a single tip location, $P_{ij}(\nu_k)$ showing dominant low-frequency noise and minor system noise peaks that are independent of $V_{\mathrm{tip}}$.
  (b) A whitened PSD for the same tip location, showing four clear frequency peaks corresponding to the fundamental frequencies of the four distinct defects identified in this dataset. By measuring the frequency drift of each of these peaks over the scan area we determine that no two exhibit a frequency that possesses a common integer factor with another at all tip locations, and thus they are not harmonics. Through triangulation over multiple scan locations or observation of the spectrogram ridge maxima for this dataset, we are further able to locate the defect origins for each signal, demonstrating that they are spatially distinct.
  }
\label{fig:SIpixel_psds}
\end{figure*}

For each scan position ($x_i,y_j$), we record $S_{21}(x_i,y_j,V_\mathrm{tip})$ while sweeping $V_{\mathrm{tip}}$ uniformly over $N_\mathrm{V}$ points with voltage step $\updelta V$, such that $V_n = V_0 + n\,\updelta V$ for $n=0,\ldots,N_\mathrm{V}-1$.
We analyse a single real quadrature obtained via a constant phase rotation,
\begin{equation}
	s_{ij}(V_{\mathrm{tip}}) \equiv \mathrm{Re} [S_{21}(x_i,y_j,V_{\mathrm{tip}})e^{-i\phi}],
\end{equation}
where $\phi$ is chosen to maximise defect contrast in the real component, $s_{ij}$, and thus may include contributions from defect-induced resonator loss and/or frequency shift.
$S_{21}$ is dominated by a large, slowly varying capacitive background that is only weakly dependent on $V_{\mathrm{tip}}$.
As per \textcite{Hegedus2024}, we isolate the fast tip--defect response by subtracting a best-fit quadratic polynomial $p^{(2)}_{ij}$,
\begin{equation}
	s_{ij,\mathrm{dtr}}(V_n)\equiv s_{ij}(V_n)-p^{(2)}_{ij}(V_n),
\end{equation}
which removes the constant offset and slow variations (e.g. due to thermal drift or electrostatic forces caused by the small tip--sample distance) whilst retaining any periodic tip--defect response.

We then compute the discrete Fourier transform of each detrended trace along $V_\mathrm{tip}$, applying a Hamming window $w_n$ as a compromise between suppressing spectral leakage and retaining sufficient frequency resolution
\begin{equation}
	\tilde s_{ij,\mathrm{dtr}}(\nu_k)=\sum_{n=0}^{N_\mathrm{V}-1} w_n\,s_{ij,\mathrm{dtr}}(V_n)\,e^{-i2\pi \nu_k V_n},
	\quad
	\nu_k=\frac{k}{N_\mathrm{V}\,\updelta V},
\end{equation}
where $k$ is the frequency bin index (thus, $\nu_k$ is the associated frequency bin).
We analyse only positive frequencies, excluding the DC bin ($k\ge 1$).
Consequently, a defect with voltage period $\Delta V_{ij}$ will fall within the bin interval
\begin{equation}
    \frac{1}{\Delta V_{ij}} \in \left[ \nu_{k-1} + \frac{1}{2 N_\mathrm{V}\,\updelta V},\ \nu_{k+1} - \frac{1}{2 N_\mathrm{V}\,\updelta V} \right).
\end{equation} 

For feature identification we use the per-pixel power spectrum
\begin{equation}
	P_{ij}(\nu_k)\equiv\left|\tilde s_{ij,\mathrm{dtr}}(\nu_k)\right|^2.
\end{equation}

To normalise and compare the tip--defect response across frequencies, we estimate a baseline $P_{\mathrm{noise}}(\nu_k)$ for each frequency bin by taking the interquartile mean of $P_{ij}(\nu_k)$ over every tip voltage sweep location $(i,j)$.
Because the tip--defect response is distance-dependent, the defect frequencies drift across the scan grid, meaning that any tip--defect response, given sufficient voltage range and sampling, will only be present in the same frequency bins over a limited set of tip locations approximately equidistant from the defect, plus any frequency crossover from other defects.
$P_{\mathrm{noise}}(\nu_k)$ therefore serves as an empirical baseline for the frequency-dependent background PSD, capturing spatially-invariant background contributions, such as temporal narrowband and broadband system noise associated with the constant $V_{\mathrm{tip}}$ slew rate, e.g. due to pulse tube cooler vibrations, as well as $V_{\mathrm{tip}}$-dependent noise.
It is also plausible that the apparent $V_\mathrm{tip}$-dependent `noise' contains a contribution of mixed signals from an ensemble of far-off Sisyphus defects that are individually too weak to resolve, however here we lump this potential contribution in with all other noise sources.
We define the whitened signal-to-noise ratio
\begin{equation}
	P^{\mathrm{w}}_{ij}(\nu_k)\equiv\frac{P_{ij}(\nu_k)}{P_{\mathrm{noise}}(\nu_k)},
\end{equation}
or in decibels
\begin{equation}
	P^{\mathrm{w,dB}}_{ij}(\nu_k)\equiv 10\log_{10}P^{\mathrm{w}}_{ij}(\nu_k).
\end{equation}
This normalisation equalises the spectral noise floor, improving the visibility of weaker periodic features across the full frequency range.
This should not, however, be used for direct comparison of the physical amplitudes of different defects, where $P_{ij}(\nu_k)$ should be used together with the local noise floor.
A comparison of non-whitened and whitened PSDs for a single tip location is provided in \autoref{fig:SIpixel_psds} with identified defects highlighted.

\begin{figure*}[htbp]
  \centering

\includegraphics[width=0.99\textwidth]{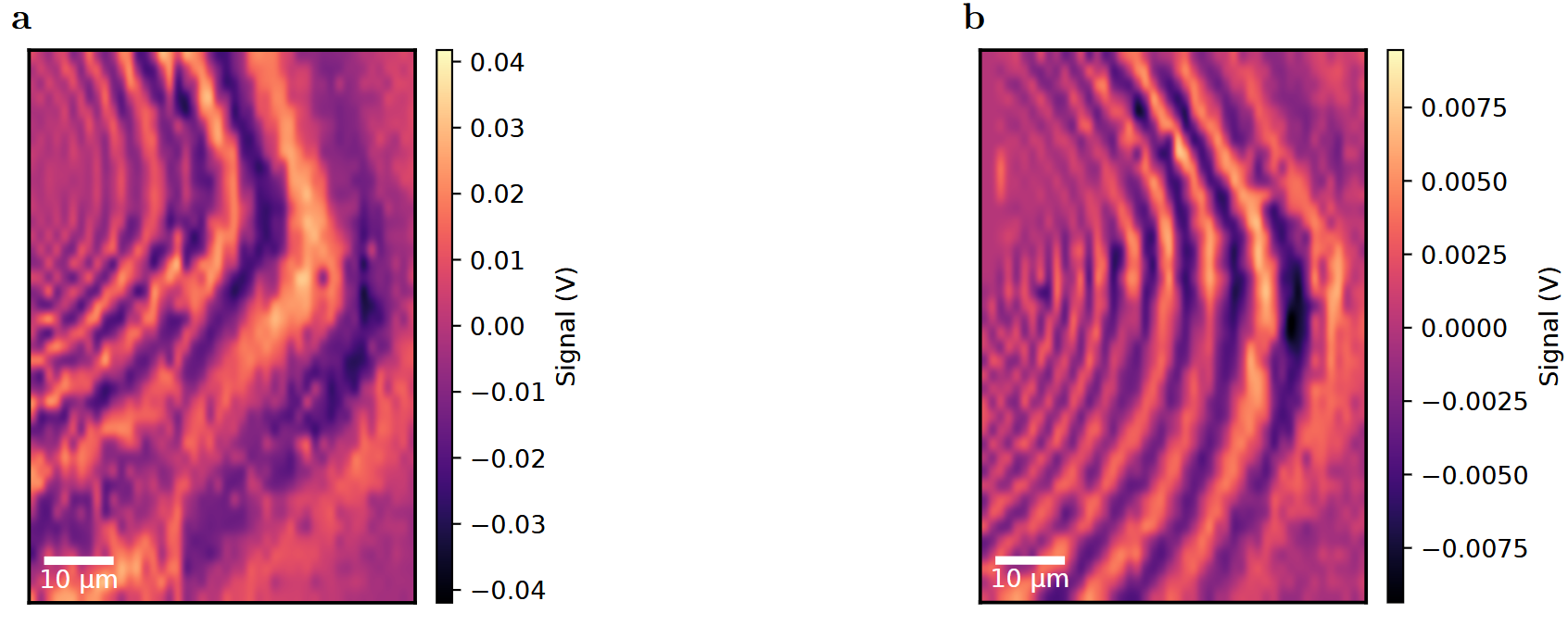}

  \caption{Fourier-extracted ridges, encompassing peaks 3 and 4 of Supplementary Figure 14, over the same scan subregion at $V_{\mathrm{tip}}=-5.491$ V, visually demonstrating their distinct origins (both to the left of the scan area). The relative signal strength of the isolated (fundamental) frequency for each defect is provided, with colour ranges normalised to maximise contrast.}
  \label{fig:SIextracted_components}
\end{figure*}

Distinct defects are identified as ridge families (comprised of a fundamental frequency ridge and sometimes also visible harmonic ridges) in spatial cross-sections (viz. with either $x$ or $y$ fixed) of $P^\mathrm{w,dB}_{ij}(\nu_k)$ that are visible above the noise floor.
Constructing a Fourier spectrogram (as in Fig.~4 of the main paper or \autoref{fig:SIresB_spectrogram}) by combining spectra along one scan axis then reveals individual Sisyphus defects as distinct ridge families sharing a common spatial centre, with peaks in frequency ($V^{-1}$) when the tip is closest to the defect.
All ridges comprising a ridge family (viz. caused by the same defect) will thus share an identical peak location in all $x$- and $y$-slices.

From this we can also isolate the contribution from a selected defect, and filter the spatial grid data to only show the contribution from a particular defect.
An example of this is shown in \autoref{fig:SIextracted_components}, where two filtered components in the two panels show clear concentric rings that have clearly different spatial origin.

To isolate a given component (indexed by $m$), we specify a frequency band $\mathcal{K}_m$ (a set of bins) that contains the ridge associated with that defect.
Where two defects overlap in frequency over parts of the scan area, we additionally restrict the analysis to a spatial sub-region where the ridge of interest dominates.
Multiple such spatially restricted reconstructions can then be merged to recover the full spatial extent of a single defect's contribution.
Within each chosen band, we pick the maximum whitened bin independently at each pixel,
\begin{equation}
	k^\ast_{ij}(m)\in\mathcal{K}_m,
	\qquad
	\nu^\ast_{ij}(m)\equiv\nu_{k^\ast_{ij}(m)}.
\end{equation}

For the single-ridge visualisations shown here, we reconstruct component $m$ by retaining only the Fourier coefficient at $k^\ast_{ij}(m)$ and zeroing all other frequency bins.
As these data are constructed from a single quadrature we enforce Hermitian symmetry to ensure the reconstructed trace is real-valued, and perform the inverse Fourier transform, yielding $s^{(m)}_{ij}(V_n)$.
Whitening is used only for peak selection via $P^{\mathrm{w}}$, whereas reconstructions use the raw coefficients $\tilde s_{ij,\mathrm{dtr}}(\nu_k)$ to preserve each component's relative amplitude.

This single-ridge reconstruction for the fundamental Fourier frequency for each defect preserves the spatial phase and period of that defect, but not the full defect waveform.
Quantitative reconstruction of defect attributes, e.g. peak amplitude, FWHM, \(\Gamma\), requires the complex additive reconstruction of all defect harmonics above the noise floor.

An example of two separated components visualised for the same $V_\mathrm{tip}$ over the same scan area, attributable to distinct Sisyphus defects, is provided in \autoref{fig:SIextracted_components}.

This different origin of the two spatially separated defect centres is further evidenced through this decomposition technique by observing that the defect in \autoref{fig:SIextracted_components}a exhibited a charge offset-jump during the grid acquisition visible as a vertical line cutting across the image, a feature that is lacking for the second defect shown in \autoref{fig:SIextracted_components}b extracted from the same dataset. This shows that these defects are physically separated, and only one responds to a change in its local charge environment.

Next, using the same filtering approach, we can extract the $|r|^\beta$-dependence of a specific defect. An example of this is shown in \autoref{fig:SIfrequency_distance_relation}.
For each reconstructed Sisyphus defect fundamental we form a spatial map of its local frequency $\nu_{ij}$, which visualises how the periodicity drifts with tip distance from the defect centre.

For defects whose centre lies within the scan area, we take $(x_0,y_0)$ as the location of the maximum observed frequency
\begin{equation}
    \nu_{ij} = A\,r_{ij}^{-\beta},
\end{equation}
where $A$ is a global amplitude and with an effective tip--defect separation
\begin{equation}
    r_{ij} = \sqrt{(x_i-x_0)^2 + (y_j-y_0)^2 + Z_{\mathrm{tip}}^2}.
\end{equation}
Here $Z_\mathrm{tip}$ is fixed such that the minimum separation at the defect centre $r_0 = Z_\mathrm{tip}$, assuming a defect located at the surface.
We compute the per-pixel exponent $\beta_{ij}$ by comparing the local frequency to the frequency at the defect centre
\begin{equation}
    \beta_{ij} = \frac{\ln \nu_0 - \ln \nu_{ij}}{\ln r_{ij} - \ln r_0}.
\end{equation}
As shown for the selected defect in \autoref{fig:SIfrequency_distance_relation}, we find as expected $\beta_{ij}$ to be consistently close to unity, with slight deviations expected due to, e.g. localised field distortions, sample height gradient, and non-uniform tip geometry.
This further supports the physical origin of the identified Sisyphus defects being islands located near the surface.

\begin{figure*}[htbp]
  \centering
\includegraphics[width=0.99\textwidth]{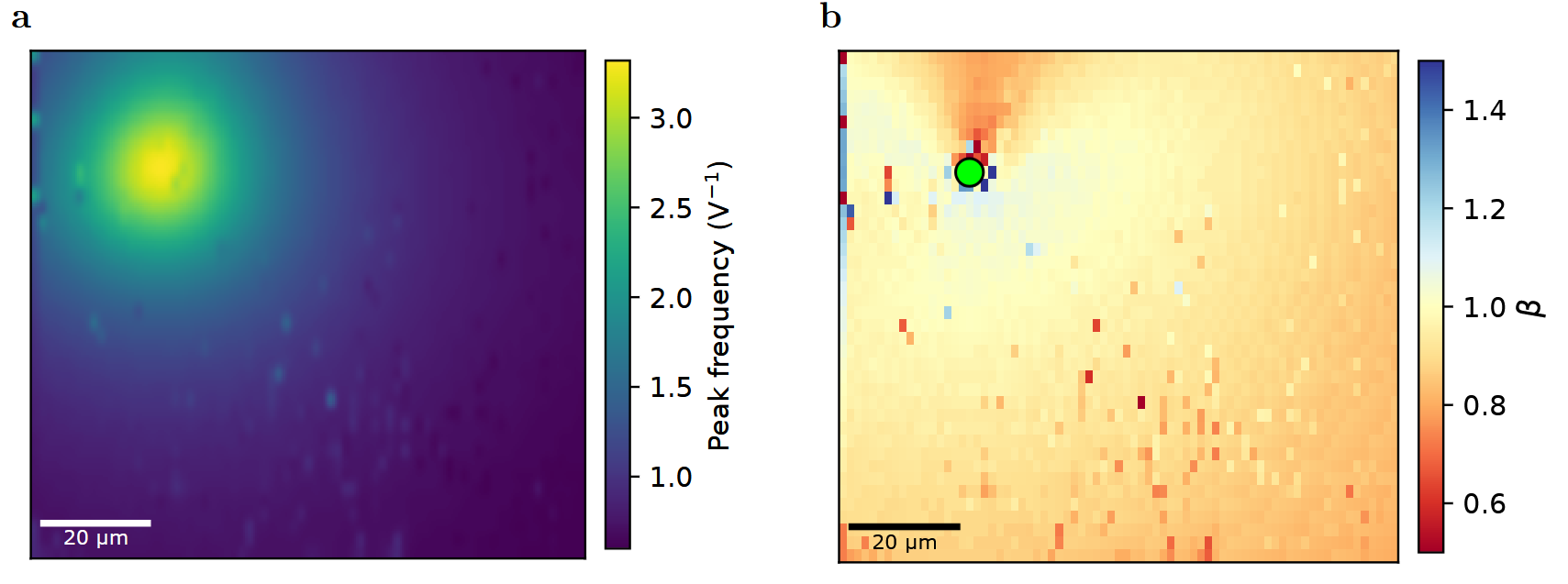}
  \caption{
  (a) shows a local frequency map for defect 3.
  (b) A per-pixel map of the required $\beta$ for the frequency--distance relation $\nu\propto r^{-\beta}$ relative to the scan location with the maximum frequency (denoted by a green dot), which we take to be the defect centre. The $\beta\approx 1$ relation (from assuming $r_0=Z_\mathrm{tip}$) present across most of the scan area is in agreement with the defect's origin being at the sample surface. 
  }
  \label{fig:SIfrequency_distance_relation}
\end{figure*}


Finally, by assuming $\beta = 1$, we may use $r = A/\nu_{ij}$, with $S_{21}$ recorded at at least three tip locations, to triangulate a defect's location within the given uncertainties previously discussed.
For each location, we evaluate
\begin{equation}
    (x_k-x_0)^2 + (y_k-y_0)^2 + Z_\mathrm{tip}^2 = \left(\frac{A}{\nu_k}\right)^2,\quad k=1,2,3
\end{equation}
which may thus be solved for the defect centre $(x_0,y_0)$ and amplitude $A$.
In practice, fitting to a larger number of data points will yield better accuracy and may be required to discern multiple defects.

\subsection{Fourier decomposition on resonator B}
The previous Fourier analysis and discussion was applied to the dataset from Resonator C. Here we finally present the defect analysis of Resonator B on the same sample, where in a single dataset we discovered at least two more Sisyphus defects. The results are shown in \autoref{fig:resonatorB_afm_spectrogram}. Here an AFM scan was taken prior to conducting the SGM measurement, and in the AFM scan we can clearly see features of the coplanar waveguide resonator, with nearby markings allowing us to identify the exact location in the resonator design. Again, the location of the defects correlate with close proximity to the patterned edges of the resonator, where we expect to be most sensitive to defects. 

\begin{figure*}[htbp]
  \centering
\includegraphics[width=0.99\textwidth]{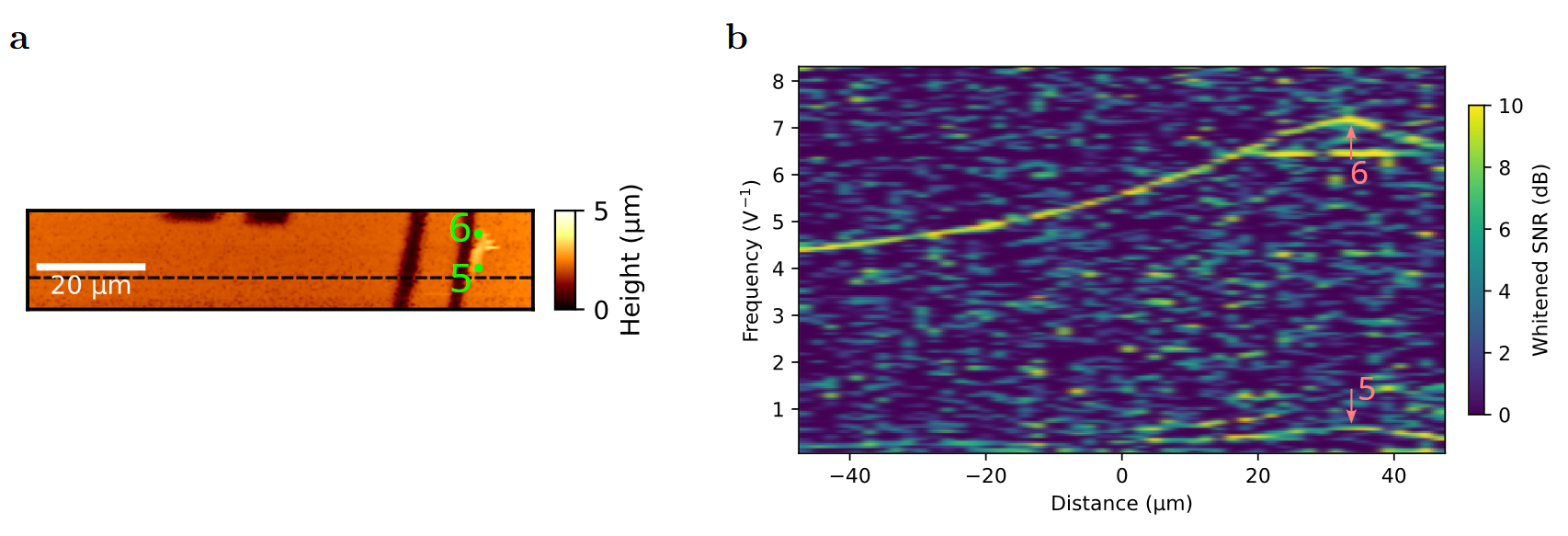}
  \caption{(a) An AFM image taken at 350 mK on the live resonator B indicating the approximate locations of two defects identified via spectrogram slices of the same region scanned using SGM. The dashed line indicates the location of the data slice in (b), which, along with complementary vertical slices (not shown) reveals both defects to have distinct origins near the edge of the resonator.} 
  \label{fig:resonatorB_afm_spectrogram}
\end{figure*}

\ifdefined\arxivversion
\else
\bibliography{main}

\end{document}
\fi

\fi

\end{document}